\documentclass[a4paper,11pt]{article}
\usepackage{aaskaiid}
\usepackage{multirow}
\usepackage{graphicx}
\usepackage{orcidlink}
\title{The 10-15 GHz radio continuum survey of the Galactic Plane with SKAO}
\ShortTitle{The 10-15 GHz radio continuum survey of the GP with SKAO}

\author[1]{A. Traficante\orcidlink{0000-0003-1665-6402}}
\ShortName{A. Traficante et al.} % shortened name list for header 
\author[1]{C. Mininni\orcidlink{0000-0002-2974-4703}}
\author[2]{F. Cavallaro\,\orcidlink{0000-0003-1856-6806}}

\author[2]{G. Umana\orcidlink{0000-0002-6972-8388}}
\author[2]{C. Trigilio}
\author[1]{S. Molinari}

\author[3]{L. D. Anderson\,\orcidlink{0000-0001-8800-1793}}
\author[4]{M. Audard\orcidlink{0000-0003-4721-034X}}
\author[2]{C. Bordiu\orcidlink{0000-0002-7703-0692}}
\author[2]{C. S. Buemi\orcidlink{0000-0002-7288-4613}}
\author[5]{C. Carrasco-Gonzalez}
\author[6]{L. Cerrigone}
\author[7]{E. J. Chung\,\orcidlink{0000-0003-0014-1527}}
\author[8]{J. Dey\,\orcidlink{0000-0003-4074-4365}}
\author[2]{A. Ingallinera\,\orcidlink{0000-0002-3137-473X}}
\author[9]{I. Jimenez-Serra\,\orcidlink{0000-0003-4493-8714}}
\author[10]{P. Klaassen\orcidlink{0000-0001-9443-0463}}
\author[2]{S. Loru}
\author[11]{K. Mallick\orcidlink{0000-0002-3873-6449}}
\author[1, 12]{A. Nucara\orcidlink{0009-0005-9192-5491}}
\author[13]{M. Padovani\orcidlink{0000-0003-2303-0096}} 
\author[14]{J. D. Pandian\orcidlink{0000-0003-4031-1121}}
\author[15]{K.~L.~J. Rygl\,\orcidlink{0000-0003-4146-9043}}
\author[16]{T. M. Rodr\'iguez\,\orcidlink{0000-0003-0090-9137}}
\author[13]{G. Sabatini\orcidlink{0000-0002-6428-9806}}
\author[17]{S. Reissl}
\author[18]{P. Suin\orcidlink{0000-0001-7044-3809}}
\author[19, 2]{M. A. Thompson\orcidlink{0000-0001-5392-909X}}
\author[20, 21]{T. L. Bourke\orcidlink{0000-0001-7491-0048}}
\author[22]{J.\,S.\,Urquhart\orcidlink{0000-0002-1605-8050}}
\author[23, 24]{M. Valeille-Manet}
\author[25]{F. Xu}
\author[26]{A. Zavagno\orcidlink{0000-0001-9509-7316}} 

\author[1]{M. Benedettini\orcidlink{0000-0002-3597-7263}}
\author[13, 24]{E. Bianchi\orcidlink{0000-0001-9249-7082}}
\author[13, 27, 28]{A. Bracco\,\orcidlink{0000-0003-0932-3140}}
\author[2]{F. Bufano\,\orcidlink{0000-0002-3429-2481}}
\author[13]{C. Codella\orcidlink{0000-0003-1514-3074}}
\author[20]{N. Cunningham\orcidlink{0000-0003-3152-8564}}
\author[29]{J. Dawson}
\author[13]{D. Galli\orcidlink{0000-0001-7706-6049}}
\author[30]{M. G. Guarcello\orcidlink {0000-0002-3010-2310}}
\author[31, 32]{A. Karska \orcidlink{0000-0001-8913-925X}}
\author[31, 33]{W.-J. Kim\,\orcidlink{0000-0003-0364-6715}}
\author[2]{P. Leto\orcidlink{0000-0003-4864-2806}}
\author[34]{B. Liu}
\author[8]{B. Mookerjea\orcidlink{0000-0003-1766-6303}}
\author[24]{F. Motte\,\orcidlink{0000-0003-1649-8002}}
\author[13, 23]{T. Nony\orcidlink{0000-0003-3246-0821}}
\author[35]{R. Paladini}
\author[22]{A. Patel}
\author[13]{L. Podio\orcidlink{0000-0003-2733-5372}}
\author[36, 37]{A.\,J.\,T. Ramaila\orcidlink{0000-0002-8167-2073}}
\author[33]{B. Riaz}
\author[2]{S. Riggi}
\author[38, 39]{D. A. Roshi\orcidlink{0000-0002-1732-5990}}
\author[2]{A. Ruggeri}
\author[40, 41]{\'A. S\'anchez-Monge\orcidlink{0000-0002-3078-9482}}
\author[42]{R. Sch{\"o}del\orcidlink{0000-0001-5404-797X}}
\author[36, 37, 15]{O.~M. Smirnov\orcidlink{0000-0003-1680-7936}}
\author[1, 43]{J.~D. Soler\orcidlink{0000-0002-0294-4465}}
\author[37]{S. Sottie\orcidlink{0000-0003-1386-023X}}
\author[44]{R. Unnikrishnan\orcidlink{0000-0002-2843-2476}}
\author[45]{A. Y. Yang\,\orcidlink{0000-0003-4546-2623}}
\author[46]{K. Wang}
\author[31]{T. Wilson}

\affiliation[1]{INAF-IAPS, Via Fosso del Cavaliere, 100, 00133 Rome, Italy}
\emailAdd{alessio.traficante@inaf.it}
\affiliation[2]{INAF - Osservatorio Astrofisico di Catania, Via Santa Sofia 78, 95123 Catania, Italy}
\affiliation[3]{Department of Physics and Astronomy, West Virginia University, Morgantown WV 26506, USA}
\affiliation[4]{Department of Astronomy, University of Geneva, Ch. Pegasi 51, 1290 Versoix, Switzerland}
\affiliation[5]{Instituto de Radioastronomía y Astrofísica (IRyA-UNAM), A. Postal 3-72 (Xangari), 58089 Morelia, Michoacán, Mexico}
\affiliation[6]{Joint ALMA Observatory, Alonso de Córdova 3107, Vitacura, Santiago 7630355, Chile}
\affiliation[7]{Korea Astronomy and Space Science Institute (KASI), 776 Daedeokdae-ro, Yuseong-gu, Daejeon 34055, Republic of Korea}
\affiliation[8]{Department of Astronomy \& Astrophysics, Tata Institute of Fundamental Research, Mumbai, 400005, India}
\affiliation[9]{Centro de Astrobiología (CAB), INTA-CSIC, Carretera de Ajalvir km 4, Torrejón de Ardoz, 28850 Madrid, Spain}
\affiliation[10]{UK Astronomy Technology Centre, Royal Observatory Edinburgh, Blackford Hill, Edinburgh EH9 3HJ, UK}
\affiliation[11]{National Astronomical Observatory of Japan, Osawa 2-21-1, Mitaka, Tokyo 181-8588, Japan}
\affiliation[12]{Dipartimento di Fisica, Università di Roma Tor Vergata, Via della Ricerca Scientifica 1, I-00133 Roma, Italy}
\affiliation[13]{INAF-Osservatorio Astrofisico di Arcetri, Largo E. Fermi 5, 50125 Firenze, Italy}
\affiliation[14]{Department of Earth \& Space Sciences, Indian Institute of Space Science and Technology (IIST), Trivandrum 695 547, India}
\affiliation[15]{INAF-IRA, Via P. Gobetti 101, 40129, Bologna, Italy}
\affiliation[16]{Physikalisches Institut, Universität zu Köln, Zülpicher Straße 77, 50937, Cologne, Germany}
\affiliation[17]{Zentrum für Astronomie der Universität Heidelberg, Institut für Theoretische Astrophysik, Albert-Ueberle-Str. 2, 69120 Heidelberg, Germany}
\affiliation[18]{Université Paris-Saclay, Université Paris Cité, CEA, CNRS, AIM, 91191 Gif-sur-Yvette, France}
\affiliation[19]{School of Physics and Astronomy, University of Leeds, Leeds LS2 9JT, UK}
\affiliation[20]{SKA Observatory, Jodrell Bank, Lower Withington, Macclesfield, SK11 9FT, UK}
\affiliation[21]{Jodrell Bank Centre for Astrophysics, School of Physics and Astronomy, University of Manchester, Oxford Road, Manchester, M13 9PL, UK}
\affiliation[22]{Centre for Astrophysics and Planetary Science, University of Kent, Canterbury, CT2 7NH, UK}
\affiliation[23]{Laboratoire d'astrophysique de Bordeaux, Univ. Bordeaux, CNRS, B18N, allée Geoffroy Saint-Hilaire, 33615, Pessac, France}
\affiliation[24]{Univ. Grenoble Alpes, CNRS, IPAG, 38000 Grenoble, France}
\affiliation[25]{Max Planck Institute for Astronomy, Königstuhl 17, 69117 Heidelberg, Germany}
\affiliation[26]{Aix Marseille Univ, CNRS, CNES, LAM Marseille, France; Institut Universitaire de France, 1 rue Descartes, 75005 Paris, France}
\affiliation[27]{Laboratoire de Physique de l’Ecole Normale Supérieure, ENS, Université PSL, CNRS, Sorbonne Université, Université de Paris, F-75005 Paris, France}
\affiliation[28]{LUX, Observatoire de Paris, PSL Research University, CNRS, Sorbonne Université, F-75014 Paris, France}
\affiliation[29]{Florida Space Institute, University of Central Florida, Orlando, Florida 32826, USA}
\affiliation[30]{Center for Advanced Research in Science and Engineering (CARSE), University of Puerto Rico, Mayag\"uez, P.R. 00681, USA}
\affiliation[31]{CSIRO Space and Astronomy, Australia Telescope National Facility, PO Box 76, Epping NSW 1710, Australia}
\affiliation[32]{INAF – Osservatorio Astronomico di Palermo, Piazza del Parlamento 1, 90134, Palermo, Italy}
\affiliation[33]{Max-Planck-Insitut für Radioastronomie, Auf dem Hügel 69, 53121, Bonn, Germany}
\affiliation[34]{Centre for Modern Interdisciplinary Technologies, Nicolaus Copernicus University in Toruń, Wileńska 4, 87-100 Toruń, Poland}
\affiliation[35]{Physikalisches Institut, Universität zu Köln, Zülpicher Str 77, D-50937 Köln, Germany}
\affiliation[36]{ National Astronomical Observatories, Chinese Academy of Sciences, Beijing, P.R. China, 100101}
\affiliation[37]{Caltech/IPAC, 1200 E. California Boulevard, Pasadena, CA 91125, USA}
\affiliation[38]{South African Radio Astronomy Observatory (SARAO), Cape Town, WC, South Africa}
\affiliation[39]{Centre for Radio Astronomy Techniques \& Technologies (RATT), Department of Physics and Electronics, Rhodes University, Makhanda, EC, South Africa}
\affiliation[40]{Institut de Ciències de l'Espai (ICE), CSIC, Campus UAB, Carrer de Can Magrans s/n, Bellaterra, E-08193 Barcelona, Spain}
\affiliation[41]{Institut d'Estudis Espacials de Catalunya (IEEC), Castelldefels, E-08860 Barcelona, Spain}
\affiliation[42]{Instituto de Astrofísica de Andalucía (CSIC), Glorieta de la Astronomía s/n, E-18008 Granada, Spain}
\affiliation[43]{University of Vienna, Department of Astrophysics, T¨urkenschanzstrasse 17, 1180 Vienna, Austria}
\affiliation[44]{Department of Space, Earth and Environment, Chalmers University of Technology, 412 96, Gothenburg, Sweden}
\affiliation[45]{National Astronomical Observatories, Chinese Academy of Sciences: Beijing, Beijing, CN}
\affiliation[46]{Kavli Institute for Astronomy and Astrophysics, Peking University, 5 Yiheyuan Road, Haidian District, Beijing 100871, People's Republic of China}

\abstract{Star formation emerges from the complex interplay between gravity, turbulence, magnetic fields, and stellar feedback, all of which vary across spatial scales and Galactic environments. Over the past decades, extensive multiwavelength surveys of the Galactic Plane have progressively unveiled this complexity. Far-infrared and sub-millimetre surveys have identified and characterized tens of thousands of star-forming regions, revealing their mass, temperature, and evolutionary stage. Complementary molecular-line surveys, spanning several CO transitions and isotopologues, have mapped the gas kinematics from giant molecular clouds down to sub-parsec structures. The advent of interferometers such as ALMA has revolutionized this field, enabling systematic studies of gas dynamics, fragmentation, and collapse in dense clumps at scales of a few thousand astronomical units.

At the same time, mid-infrared and radio surveys at frequencies $0.8 \lesssim \nu \lesssim 5$ GHz have traced ionised gas associated with the earliest and latest phases of massive-star evolution, including thermal radio jets, hypercompact and ultracompact \hii{} regions, supernova remnants, planetary nebulae, and evolved massive stars. Yet, a uniform, Galaxy-wide census of ionised structures and feedback processes remains elusive.

A transformational leap forward requires a sensitive, high-resolution radio survey of the Galactic Plane at 10–15 GHz, capable of resolving physical scales smaller than 0.05 pc at distances up to 20 kpc. This is precisely the goal of the SKA-Mid Galactic Plane survey, which will, with its unprecedented sensitivity, angular resolution, and mapping speed, provide the first panoptic view of ionised gas and stellar feedback across the Milky Way.}

\newcommand{\msun}{\mbox{M$_\odot$}}

\newcommand{\hii}{\mbox{H\,{\sc{ii}} }}
\newcommand{\meth}{$\mathrm{CH_3OH}$}
\newcommand{\form}{$\mathrm{H_2CO}$}

\newcommand{\arcsec}{$^{\prime\prime}$}

\def\tablefoot#1{\par\vspace*{2ex}%
 \parbox{\hsize}{\leftskip0pt\rightskip0pt
 {\noindent\small\textbf{Notes.}~#1\par}}}
\def\tablefootmark#1{$^{#1}$\,\ignorespaces}
\def\tablefoottext#1#2{$^{(#1)}$~#2}
\def\tablebib#1{\par\vspace*{2ex}%
 \parbox{\hsize}{\leftskip0pt\rightskip0pt
 {\noindent\small\textbf{References.}~#1\par}}}

\begin{document}
\newcommand{\actaa}{Acta Astron.} % Acta Astronomica
\newcommand{\araa}{ARA\&A} % Annual Review of Astron and Astrophys
\newcommand{\aar}{A\&ARv} % Astrononmy \& Astrophysics Review
\newcommand{\aapr}{A\&ARv} % Astronomy\&Astrophysics Reviews
\newcommand{\ab}{Astrobiol.} % Astrobiology
\newcommand{\aj}{AJ} % Astronomical Journal
\newcommand{\apj}{ApJ} % Astrophysical Journal
\newcommand{\apjl}{ApJL} % Astrophysical Journal, Letters
\newcommand{\apjs}{ApJSS} % Astrophysical Journal, Supplement
\newcommand{\ao}{Appl. Opt.} % Applied Optics
\newcommand{\apss}{Astro. \& Space Sci.} % Astrophysics and Space Science
\newcommand{\aap}{A\&A} % Astronomy and Astrophysics
\newcommand{\aaps}{A\&AS.} % Astronomy and Astrophysics, Supplement
\newcommand{\baas}{Bull. Am. Astron. Soc.} % Bulletin of the AAS
\newcommand{\caa}{Chinese A\&A} % Chinese Astronomy and Astrophysics
\newcommand{\cjaa}{Chinese J. A\&A} % Chinese Journal of Astronomy and Astrophysics
\newcommand{\cqg}{Class. Quantum Gravity} % Classical and Quantum Gravity
\newcommand{\gal}{Galaxies} % Galaxies
\newcommand{\gca}{Geo. Cosmo. Acta} % Geochimica Cosmochimica Acta
\newcommand{\icarus}{Icarus} % Icarus
\newcommand{\jcap}{JCAP} % Journal of Cosmology and Astroparticle Physics
\newcommand{\jgr}{J. Geophys. Res.} % Journal of Geophysics Research
\newcommand{\jgrp}{J. Geophys. Res. Planets} % Journal of Geophysics Research: Planets
\newcommand{\jqsrt}{J. Quant. Spectrosc. Radiat. Transf.} % Journal of Quantitiative Spectroscopy and Radiative Transfer
\newcommand{\memsai}{Mem. SAIt} % Mem. Societa Astronomica Italiana
\newcommand{\mnras}{MNRAS} % Monthly Notices of the RAS
\newcommand{\nat}{Nature} % Nature
\newcommand{\nastro}{Nat. Astron.} % Nature Astronomy
\newcommand{\ncomms}{Nat. Commun.} % Nature Communications
\newcommand{\nphys}{Nat. Phys.} % Nature Physics
\newcommand{\na}{New Astron.} % New Astronomy
\newcommand{\nar}{New Astron. Rev.} % New Astronomy Review
\newcommand{\physrep}{Phys. Rep.} % Physics Reports
\newcommand{\pra}{Phys. Rev. A} % Physical Review A: General Physics
\newcommand{\prb}{Phys. Rev. B} % Physical Review B: Solid State
\newcommand{\prc}{Phys. Rev. C} % Physical Review C
\newcommand{\prd}{Phys. Rev. D} % Physical Review D
\newcommand{\pre}{Phys. Rev. E} % Physical Review E
\newcommand{\prl}{Phys. Rev. L.} % Physical Review Letters
\newcommand{\psj}{Planet. Sci. J.} % Planetary Science Journal
\newcommand{\planss}{Planet. Space Sci.} % Planetary Space Science
\newcommand{\pnas}{Proc. Natl Acad. Sci. USA} % Proceedings of the US National Academy of Sciences
\newcommand{\procspie}{Proc. SPIE} % Proceedings of the SPIE
\newcommand{\pasa}{PASA} % Publications of the Astron.  Soc. of Australia
\newcommand{\pasj}{PASJ} % Publications of the Astron.  Soc. of Japan 
\newcommand{\pasp}{PASP} % Publications of the Astron.  Soc. of the Pacific
\newcommand{\rmxaa}{RMXAA} % Revista Mexicana de Astronomia y Astrofisica
\newcommand{\sci}{Science} % Science
\newcommand{\sciadv}{Sci. Adv.} % Science Advances
\newcommand{\solphys}{Sol. Phys.} % Solar Physics
\newcommand{\sovast}{Soviet Ast.} % Soviet Astronomy
\newcommand{\ssr}{Space Sci. Rev.} % Space Science Reviews
\newcommand{\uni}{Universe} % Universe

\maketitle

%%%%%%%%%%%%%%%%%%%%%%%%%%%%%%%%%%%%%%%%%%%%%%%%%%%%%%%%%%%%%%%%%%%%%%%%%%%%%%%%%%
\section{Introduction}
Star formation is the outcome of a complex, hierarchical process that operates across multiple spatial scales — from hundreds of parsecs-scale molecular clouds down to parsec-scale clumps and thousands of AU dense cores \citep{Motte18, Beuther25}. At each scale, the interplay of gravity, turbulence, magnetic fields, and feedback determines the morphology, fragmentation, and dynamical evolution of the gas \citep{Vazquez-Semadeni19, Vazquez-Semadeni25, Padoan20, Traficante18, Traficante20, Peretto23}. Understanding the interplay between these processes across various spatial scales is crucial for resolving a decades-old enigma in star formation theory:  what regulates the star formation rate (SFR) in our Galaxy. The total mass of molecular hydrogen (H$_{2}$) in Galactic molecular clouds is estimated to be $\simeq10^{9}$ M$_{\odot}$ \citep{Miville-Deschenes17}. If this gas was fully gravitationally bound and allowed to undergo global gravitational collapse, it would do so on a free-fall timescale of roughly $\simeq10^{7}$ years. This scenario would imply a Galactic SFR of about $200$ M$_{\odot}$/yr. However, the SFR stands among the most precisely constrained quantities in astrophysics, with concordant determinations spanning tracers from the cosmic microwave background to the far-infrared and X-ray regimes, all converging at a value near $2$ M$_{\odot}$/yr \cite[][and references therein]{Elia22}. 

While gravity remains the fundamental driver of collapse across all spatial scales, promoting the formation of filamentary networks and dense hubs \citep{Kumar20, Hacar23}, supersonic turbulence can either inhibit or promote this collapse by generating local overdensities within the gas. Magnetic fields further modulate this interplay by regulating angular momentum transport and influencing filament alignment \citep[][]{Pattle23}. Crucially, both small- and large-scale feedback processes inject energy and momentum into the interstellar medium (ISM), thereby shaping its evolution. In particular, feedback from massive stars, through mechanisms such as ionising radiation from emerging \hii regions, stellar winds, outflows, and supernova explosions strongly impact the surrounding medium \citep{Grudic22}. They all drive turbulence and influence the balance between suppression and triggering of subsequent star formation on scales of tens of parsecs from their natal sites. Understanding the impact of these feedback across different spatial scales and Galactic environments is crucial for identifying the mechanisms that substantially suppress the Galactic SFR. What all of these mechanisms (except the supernovae) have in common is that they ionise the surrounding medium and generate thermal bremsstrahlung (free-free) emission. 

The scientific motivation for conducting a Galactic Plane (GP) survey in the 10--15\,GHz range is to investigate feedback mechanisms spanning across many different Galactic environments and in a vast range of spatial scales. These high-frequency radio observations are in fact particularly sensitive to free-free emission, and are less contaminated by synchrotron emission or anomalous microwave emission than that at lower radio frequencies. Supernovae explosions generate gyrosynchrotron emission, but it is very likely that in some of these regions even at these radio frequencies this is still the dominant form of radiation. These higher radio frequencies, required to study ionized gas, remain vastly unexplored.

In this chapter, we present the scientific framework of a GP survey with the Square Kilometre Array Observatory (SKAO), discuss the key science goals, outline the main survey design requirements and highlight the transformative potential of such an effort in addressing long-standing questions on the interplay between massive stars and the ISM.

First, in Section \ref{sec:ancillary_surveys} we review the state-of-the-art of the GP surveys at various range of wavelengths: infrared and sub-mm (Section \ref{section_IR_Surveys}), CO and other molecular lines (Section \ref{section_CO_MolecularLineSurveys}), optical and Xray (Section \ref{section:optical_Xray}), mm high-resolution surveys with interferometers (Section \ref{section_SynergyWithALMA}) and radio surveys at low ($\nu\lesssim5$ GHz) frequencies (Section \ref{section_RadioContinuumSurveys}, concluding the Section with the needs for a new survey of the GP in the radio frequencies enabled by SKA-Mid in Band 5b ($\sim 8.3-15.3$ GHz, Section \ref{section:needs_SKA_survey}). In Section \ref{sec:science_with_GP} we outline the diverse range of science cases enabled by the proposed GP survey with SKA-Mid. The primary goal will be the investigation of feedback from \hii regions (Section\ref{section_Feedback_HIIRegions}), since such a survey will uniquely enable statistical studies of their physical properties, morphologies, evolutionary states, and impact on subsequent generations of star formation throughout the GP. Beyond this primary goal, the proposed survey will allow the detection and characterisation of thousands of different objects across a broad range of evolutionary stages, including radio jets from deeply embedded protostars (Section \ref{sec:radio_jets}), supernova remnants (SNRs, Section \ref{sec:supernovae}), planetary nebulae (PNe, Section \ref{sec:planetary_nebulae}) and other evolved stellar processes (Section \ref{sec:evolved_massive_star}). In Section \ref{sec:models_simulations} we discuss new models that can improve the interpretation of our data: how the contribution of synchrotron emission can still influence the free-free emission even at the band 5b frequencies (Section \ref{sec:synchrotron_free_ree}), and we introduce a novel approach to post-process and model the free-free emission seen by SKA-Mid at $\sim10-15$ GHz following the approach of the Rosetta Stone project (\citealt{Lebreuilly25,Tung25,Nucara25}, Section \ref{sec:RS_post_processing}). In Section \ref{sec:synergies} we outline the most significant synergies that the proposed radio continuum survey will provide with other projects achievable with SKAO, such as the study of star-forming regions towards radio recombination lines (Section \ref{sec:RRLs}) or methanol masers (Section \ref{sec:methanol_masers}, as well a detailed study of the most extreme region in our Galaxy, the Galactic Center (Section \ref{sec:Galactic_center}). In Section \ref{sec:technical_requirements} we present the observational strategy and the technical requirements to realize the GP survey in band 5b with SKA-Mid and finally in section \ref{sec:conclusions} we draw our conclusions and final remarks.

% In this chapter, we present the scientific framework of a GP survey with the SKA, discuss the key science goals, outline the main survey design requirements and highlight the transformative potential of such an effort in addressing long-standing questions on the interplay between massive stars and the ISM.

%%%%%%%%%%%%%%%%%%%%%%%%%%%%%%%%%%%%%%%%%%%%%%%%%%%%%%%%%%%%%%%%%%%%%%%%%%%%%%%%%%
\section{Galactic Plane Surveys: Legacy, Achievements, and Limitations}\label{sec:ancillary_surveys}
% Surveys of the Galactic Plane (GP) have long played a fundamental role in advancing our understanding of the structure, content, and evolution of the interstellar medium ISM and star formation processes in the Milky Way. Across the electromagnetic spectrum, from the near-infrared to the mm and radio domains, large-scale surveys have enabled the characterisation of tens of thousands of Galactic objects, including massive star-forming regions \citep{Benjamin03, Carey05, Schuller2009, Molinari10_PASP}, supernova remnants, compact \hii regions \citep{Anderson09, Urquhart2013}, and the more diffuse gas components such as HI, molecular complexes and ionised gas \citep{Dame01, Beuther16}. The synergy between these surveys has allowed us to advance towards a comprehensive, panoptic theory of the star formation mechanism.

Surveys of the GP have long played a fundamental role in advancing our understanding of the structure, content, and evolution of the ISM and star formation processes in the Milky Way, allowing statistically significant census of Galactic structures and their characterization in different environments. A number of large-scale surveys have been carried out in the past two decades across a broad range of wavelengths, each contributing to a progressively more complete view of the gravitational content and the turbulent gas dynamics of the various regions across the Galaxy. As discussed in detail in the following sections, these surveys span a wide range of spatial scales and sensitivities but they are nevertheless highly complementary, tracing different phases of the interstellar medium or emphasizing specific physical aspects of the same phase. These surveys include far-infrared (FIR) and sub-mm continuum surveys of the cold dust reservoirs, large-scale CO surveys tracing molecular gas kinematics, optical and X-ray surveys aimed to identify thousands of young stellar objects (YSOs) and a series of radio continuum and spectral line surveys at cm wavelengths. Together, they have enabled the identification and characterisation of tens of thousands of Galactic objects, including massive star-forming regions, SNRs, compact \hii regions and the more diffuse gas components such as HI, molecular complexes and ionised gas.

In the next sections we will summarize the major results obtained from these surveys in the past years and we highlight how the synergy between these surveys and the proposed GP survey in band 5b will finally allow us to advance towards a comprehensive, panoptic theory of the star formation process in our Galaxy. % the fundamental and complementary role that will be played by the proposed SKA GP survey. 

%%%%%%%%%%%%%%%%%%%%%%%%%%%%%%%%%%%%%%%%%%%%%%%%%%%%%%%%%%%%%%%%%%%%%%%%%%%%%%%%%%

\subsection{Infrared and sub-mm Surveys}
\label{section_IR_Surveys}
Far-infrared and sub-mm surveys have mapped the thermal dust emission associated with cold molecular gas, offering essential constraints on column density, dust temperature, and mass distribution across the GP. These surveys are pivotal in identifying the earliest phases of star formation, including pre-stellar cores and filamentary cloud structures. In particular, the Herschel Hi-GAL survey from $70-500\,\mu$m \citep{Molinari10_PASP} covered the entire Galactic plane. The $870\,\mu$m ATLASGAL \citep{Schuller2009} and $\simeq1.1$ \,mm Bolocam Galactic Plane Survey \citep[BGPS,][]{aguirre11} focused on the inner region of the Galaxy. Together, these surveys identified and characterized the physical properties of more than 30000 filamentary structures (\citealt{Li16, Schisano20}) and more than $2\times10^{5}$ star-forming parsec-scales clump \citep{Contreras2013, Csengeri14, Svoboda16, Elia21} across the Galaxy.

While they do not trace ionised gas directly, they are essential for placing \hii regions and other feedback-driven structures within the broader molecular cloud environment. For example, they allowed for a detailed evolutionary classification of star-forming clumps across the Galaxy based on the widely used luminosity over mass (L/M) indicator \citep{Molinari08, Molinari16, Duarte-Cabral13, Merello19, Traficante18b, urquhart2022} and in synergy with radio observations have been crucial to recognize that clumps with L/M$\geq10$ are the best candidates to be associated with newly formed \hii regions \citep{urquhart2013_cornish,Cesaroni15}.

There have been a large number of GP surveys also in the mid-infrared. The most relevant for characterising the star formation sites are the {\textit{Spitzer}} surveys: the Galactic Legacy Infrared Mid-Plane Survey Extraordinaire \citep[GLIMPSE,][]{Benjamin_GLIMPSE_2003PASP} from $3-9\,\mu$m and the MIPSGAL survey at $24\,\mu$m \citep{Carey_MIPSGAL_2009PASP}. Together, these surveys have allowed the identification and classification of thousands of young star forming clouds seen in absorption against the mid-infrared background, the so-called infrared dark clouds (IRDCs, \citealt{Peretto09}). And, in combination with the Hi-GAL survey we were able for the first time to identify, classify and characterize thousands of young, massive clumps associated with IRDCs \citep{Traficante15}.

% identified hundreds of thousands of newly formed protostars and together with the FIR-sub mm surveys they have allowed a comprehensive classification of the evolution of star-forming clumps.

At these wavelengths, surveys of the GP have also uncovered a significant number of young stellar objects with outflow candidates and \hii regions. For example, more than 300 outflow candidates have been identified thanks to the extended emission in the 4.5 $\mu$m GLIMPSE maps (the so-called Extendend Green Objects; \citealt{Cyganowski08}). The GLIMPSE survey has also been fundamental to identify $\sim600$ bubbles candidate to be formed by newly born \hii regions \citep{Churchwell_Bubbles_2006ApJ,Churchwell_Bubbles_2007ApJ,Churchwell_2009PASP}.

The Wide-field Infrared Survey Explorer (WISE) telescope in the mid-IR regime has also yielded a notable sample of \hii regions. Thanks to its full-sky survey, a deep investigation of the emission at 12 $\mu$m and 22 $\mu$m allowed the identification of $\simeq8000$ \hii and candidate \hii regions across the GP \citep{Anderson_WiseHII_2014ApJS}. This has been mainly possible through the lens of PAH emission features, which are excited by the UV radiation field of massive stars in photodissociation regions \citep[PDR,][]{Tielens_PDRs_2023ASSP}, and due to the emission by heated dust at a few tens of microns. While PAH emissions demarcate the boundary of the PDR, radio continuum observations will help in delineating the ionization content in a given region, with the presence of heated dust emission within the front as an added indicator. The synergy between IR and radio surveys will help to understand morphological structures and the strength of the feedback mechanisms, which are often needed to constrain star formation scenarios of different complex regions.

\subsection{CO and Molecular Line Surveys}
\label{section_CO_MolecularLineSurveys}

Extensive surveys of CO and their isotopologues (e.g., $^{13}$CO, C$^{18}$O), as well as higher-density tracers such as HCN and HCO$^+$, have been conducted across large portions of the GP using various facilities, providing insights into the distribution, kinematics, and turbulent properties of molecular gas. The CfA 1.2~m survey provided the first comprehensive map of $^{12}$CO (1–0) emission across the entire Milky Way disk, laying the foundation for subsequent studies \citep{dame2001}. The FUGIN (FOREST Unbiased Galactic plane Imaging survey with the Nobeyama 45m telescope) survey, using the Nobeyama 45~m telescope, delivered high-resolution maps of $^{12}$CO, $^{13}$CO, and C$^{18}$O (1–0) lines over the inner Galaxy \citep{umemoto2017}. The Galactic Ring Survey (GRS) focused on $^{13}$CO (1–0) to study the molecular ring structure \citep{jackson2006}, while the Mopra CO survey extended such efforts into the southern plane with observations of $^{12}$CO, $^{13}$CO, and C$^{18}$O J=1–0 lines \citep{burton2013}. Similar longitudes, but extended latitude ranges have been covered in multiple CO $(1-0)$ isotopologues by the Three-mm Ultimate Mopra Milky Way Survey (ThRUMMS, \citealt{Barnes15}). Higher-excitation transitions have been targeted by CHIMPS (CO Heterodyne Inner Milky Way Plane Survey; $^{13}$CO and C$^{18}$O J=3–2) and COHRS (CO High-Resolution Survey; $^{12}$CO J=3–2) using the JCMT, revealing the distribution of warmer and denser gas \citep{rigby2015,dempsey2013}. The SEDIGISM (Structure, Excitation, and Dynamics of the Inner Galactic Interstellar Medium) survey complements these by mapping $^{13}$CO and C$^{18}$O J=2–1 emission over a wide longitude range with the APEX telescope, enabling detailed studies of the physical and dynamical structure of the inner Galactic disk \citep{schuller2017,Schuller2021}. The Outer Galaxy High-Resolution Survey (OGHReS; \citealt{colombo2021, urquhart2024_oghres, urquhart2025}) is a highly complementary APEX CO survey of the 3rd quadrant outer Galaxy ($180^\circ < \ell < 280^\circ$) allowing the study of molecular clouds and star formation to lower density, pressure and metallicity environments. The Forgotten Quadrant Survey \citep[FQS,][]{Benedettini20, Benedettini21} has mapped a portion of the 2nd quadrant ($220^\circ < \ell < 240^\circ$) and $-2.5^\circ < b < 0^\circ$) in both $^{12}$CO (1–0) and $^{13}$CO (1–0) to identify and to study the kinematics of $\simeq250$ molecular clouds in this poorly explored region of our Galaxy.

\begin{figure*}
    \centering
	\includegraphics[width=1\columnwidth]{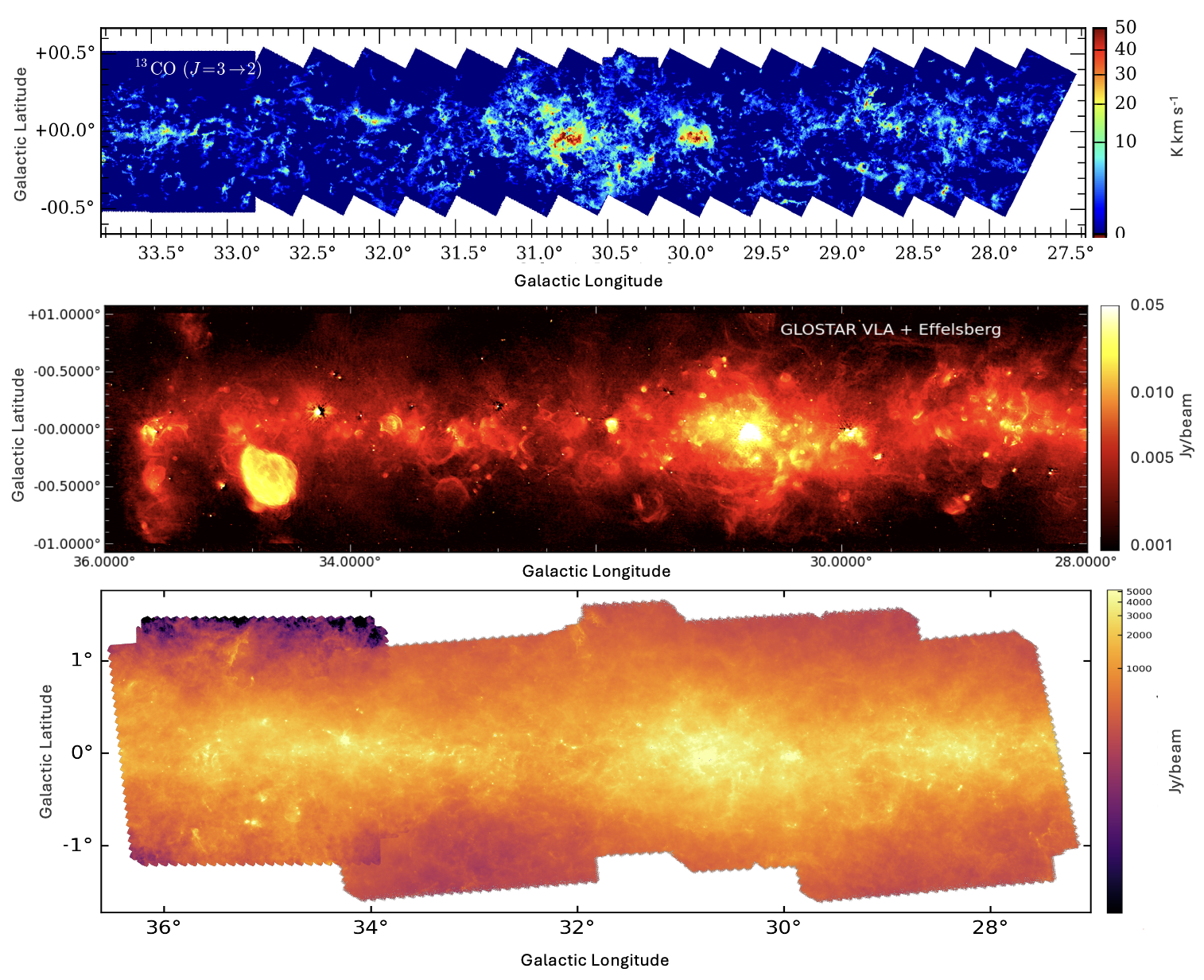}
    \caption{Example of GP surveys at different wavelengths. \textit{Top panel}: $^{13}$CO (3$-$2) emission image from CHIMPS (adapted from \citealt{rigby2015}). \textit{Middle panel}: 5~GHz image from GLOSTAR (adapted from \citealt{Brunthaler2021}). \textit{Bottom panel}: Hi-GAL 250 $\mu$m emission (adapted from \citealt{Traficante11} and \citealt{Molinari10_PASP}).}
    \label{fig:example_figure}
\end{figure*}

Complementing the CO-based surveys, targeted observations of dense gas tracers have also been conducted across the GP. The most extensive among them is the MALT90 (Millimeter Astronomy Legacy Team 90 GHz, \citealt{jackson2013}) survey, which used the Mopra 22-m telescope to observe 16 molecular transitions -- including HCN(1–0) and HCO$^{+}$ (1–0) (\citealt{rathborne2016}) -- toward over 2000 dense clumps identified by the ATLASGAL submillimeter continuum survey \citep{Contreras2013, Urquhart2014}. The CHaMP (Census of High- and Medium-mass Protostars) survey also employed Mopra to map HCO$^{+}$ (1–0) emission across a section of the southern GP ($\ell \sim 280^{\circ} - 300^{\circ}$), focusing on high- and intermediate-mass star-forming regions \citep{barnes2011}. More recently, the JCMT’s MAJORS (Massive, Active, JCMT-Observed Regions of Star formation) program has begun targeting HCN (3–2) and HCO$^{+}$ (3–2) emission toward a mass-selected sample of star-forming regions (\citealt{urquhart2014_rms}), including sources in the GP and the Central Molecular Zone (Eden et al., in preparation).

All these surveys are essential for determining the physical conditions of the molecular clouds that give rise to massive stars \citep{Miville-Deschenes17}, and for constraining the gravitational and turbulent balance of star-forming clumps \citep{Traficante18b}, filaments and clouds \citep{Duarte-Cabral21}. However, they do not directly trace the ionised gas or its spatial connection with feedback-related structures.

% \begin{itemize}
%     \item Contributor : \textcolor{teal}{E. J. Chung}, \textcolor{purple}{J. Urquhart}
% \end{itemize}

%%%%%%%%%%%%%%%%%%%%%%%%%%%%%%%%%%%%%%%%%%%%%%%%%%%%%%%%%%%%%%%%%%%%%%%%%%%%%%%%%%
\subsection{Optical / X-ray Surveys}
\label{section:optical_Xray}
% \textcolor{
Optical and X-ray surveys give an opportunity to identify objects like YSOs and flaring stars (and in general, stars) in our solar neighborhood and beyond. The former helped to reveal more evolved YSOs, while X-ray surveys are efficient at detecting embedded sources and, in particular, at disentangling foreground stars from younger pre-main sequence objects. Observations at these wavelengths are providing a complementary view of physical mechanisms in stars detected in a GP survey, such as  magnetic activity, wind emission, aurorae. For example, the Gaia space observatory provides an all-sky survey for about 2 billion sources (mainly stars in our Galaxy, with about 10 million extragalactic sources; \citealt{GaiaDR3,GaiaDR3Xgal}) in the optical ($G$ band and sub-bands $BP$ and $RP$, including spectro-photometry), allowing for precise position, distance, and kinematic measurements up to several kpc, and provides mid spectral resolution spectra (and thus radial velocities) for millions of bright sources in the near-infrared (845–872 nm).   Gaia, furthermore, scans the sky repeatedly, offering a window into the time domain variability, and thus, classification of variable stars \citep[e.g., about 10 million variables in DR3, and 80k young stars across the Galaxy,][]{Eyer23,Rimoldini23,Marton23}.  The Vera Rubin Observatory (VRO/LSST) has recently started its observations and will observe repeatedly a large portion of the GP in different filters. Similarly, in the X-ray regime, all-sky surveys including the GP were done by eROSITA \citep[e.g., 230k sources in DR1 for the Western Galactic Hemisphere, i.e., Galactic longitudes $0^{\circ}-180^{\circ}$,][]{Merloni2024,Salvato2025}, ROSAT \citep[e.g., 145k sources across all sky, among them 18.8k bright X-ray sources and 28k identified as stars,][]{Voges1999,RASSstars}, and by XMM-Newton or Chandra (e.g., 427k sources detected in the 4XMM-DR14s catalogue of serendipitous sources, \citealt{Traulsen19,Traulsen20}; 408k sources detected in the Chandra Source Catalog Release 2.1, \citealt{Evans24}).

All these surveys probe the large-scale structure, gas dynamics and stellar population distribution across the Galaxy. However, they do not provide information about the future population of stars in star-forming regions. This is what we will describe in the following Section.

%%%%%%%%%%%%%%%%%%%%%%%%%%%%%%%%%%%%%%%%%%%%%%%%%%%%%%%%%%%%%%%%%%%%%%%%%%%%%%%%%%

\subsection{ALMA/NOEMA surveys of star-forming regions}
\label{section_SynergyWithALMA}

The investigation of the dense, cold star-forming regions down to thousands of AU scales have been recently made possible thanks to mm/sub-mm observations with powerful interferometers such as ALMA and NOEMA. These observations provide unprecedented spatial and spectral resolution studies of all evolutionary stages, from the cold starless clump/core stage in IRDCs up to the Hot Molecular Core and the \hii region stages, wherein the \hii region itself matures from a hypercompact to ultracompact to compact to classical form.

The earliest stages of star formation have been studied in different ALMA surveys \citep{Svoboda19, Anderson21}, including the ASHES \citep[The ALMA Survey of 70 $\mu$m Dark High-mass Clumps in Early Stages;][]{Sanhueza_ASHES_2019ApJ} program, which together have mapped tens of 70 $\mu$m massive dark clumps. At the more evolved phases, the NOEMA CORE survey \citep{Beuther18}, together with the ATOMS \citep[ALMA Three-millimeter Observations of Massive Star-forming regions;][]{Liu_ATOMSI_2020MNRAS}, QUARK \cite[Querying Underlying mechanisms of massive star formation with ALMA-Resolved gas Kinematics and Structures;][]{Xu24a}, ASSEMBLE \cite[ALMA Survey of Star Formation and Evolution in Massive Protoclusters with Blue Profiles;][]{Xu24b} or the DIHCA \cite[Digging into the Interior of Hot Cores with ALMA;][]{Ishihara24} surveys have covered hundreds of evolved star-forming regions at the hot core or UC/HC\,\hii region stages both in continuum and molecular lines and investigate the fragmentation properties and the gas dynamics in regions strongly affected by early feedback mechanisms. Other surveys have been also specifically designed to investigate the complex chemistry during the hot core stage of the massive star formation \citep[Complex Chemistry in hot Cores with ALMA (CoCCoA);][]{Chen_CoCCoA_2023AA}. Few surveys have instead mapped clumps at different evolutionary stages with the aim of investigating the evolution of the accretion rates in globally collapsing clumps \cite[Star formation in QUiescent And Luminous Objects (SQUALO);][]{Traficante23} or the fragmentation properties across bright, luminous objects \citep[Tracing Evolution in Massive Protostellar Objects (TEMPO);][]{Avison_TEMPO_2023MNRAS}.

The ALMA telescope has also allowed a fundamental step forward in our understanding of the fragmentation properties in massive star forming regions thanks to large surveys of the GP aimed to explore in a statistically significant way the fundamental role of the environment in the hierarchical, multi-scale star formation mechanism. The ALMAGAL survey \citep{Molinari_ALMAGALI_2025AA}, an ALMA large program at $\lambda\simeq1.3$ mm has mapped more than 1000 dense clumps -- spanning the full evolutionary range from the IRDCs to \hii regions -- with a minimum resolution of $\sim$1000\,AU, and provided the first characterization of the fragmentation properties at these scales on a Galaxy-wide survey \citep{Coletta25}, as well as the first characterization of the morphology of different molecular line emissions across these massive clumps at various evolutionary stages \citep{Mininni25}. The ALMA-IMF large program \citep{Motte_ALMAIMF_2022AA} has characterised
the molecular and ionized gas environment of 15 massive protoclusters on the GP, investigating the core populations in different protoclusters \citep{Pouteau22, Nony23, Armante24} as well as providing 
a sample of $\simeq1000$ gravitationally bound cores with $\sim$2000\,AU, radii and well-constrained properties: prestellar or protostellar, mass, protostellar luminosity, association with hot cores and ionized gas \citep{Bonfand24,Galvan24,Louvet24,Motte25}.

The picture that is emerging is consistent with the family of models defined as "clump-fed" scenario, in which the fragmentation is very dynamical and involves a multi-scale process where the gas from the parent parsec-scale clump (and likely the larger scales filament) participate at the collapse and formation of the inner cores \citep{Vazquez-Semadeni19, Vazquez-Semadeni25, Padoan20}. The mass of observed cores has a direct relation with that of the parent clump, and the distribution of fragments within each clump seems to evolve with the clump evolution \citep{Traficante23, Xu24a}. The question of how this gas flows at various scales to produce the observed fragmentation properties, however, remains still elusive. At what scales the dynamics is dominated by the gravitational collapse, the turbulent motions or the feedback mechanisms is still unclear. The multi-scale gas dynamics in different environments has now started to be investigated with ALMA surveys in IRDCs \citep{Traficante20, Peretto23} and in massive clusters at various phases with ALMA-IMF \citep{Sandoval-Garrido25, Koley25}. The first large survey of multi-scale gas dynamics in different Galactic environments, combining cloud-scales information down to thousands of AU core scales is now going to be carried out thanks to the ALMA Large project PANTA REI (PIs: N. Peretto; A. Traficante; S. Clark; M. Merello). This project will map the 3-mm emission associated with 259 clumps already observed at the tens of parsec scales with CO surveys such as SEDIGISM \citep{Duarte-Cabral21} or OGHReS \citep{colombo2021} and, at the few thousands of AU scales with ALMAGAL \citep{Molinari_ALMAGALI_2025AA}, filling the gap between these scales and allowing a comprehensive study of the interplay between turbulence and gravity in different Galactic environments. Furthermore, by observing molecular tracers such as SiO (2-1) or HCO$^{+}$ (1-0) PANTA REI will allow a statistically significant investigation of outflows \citep{Duarte-Cabral14, Traficante17} and their impact on the gas dynamics at all spatial scales. 

\begin{figure}[!ht]
    \includegraphics[width=\linewidth]{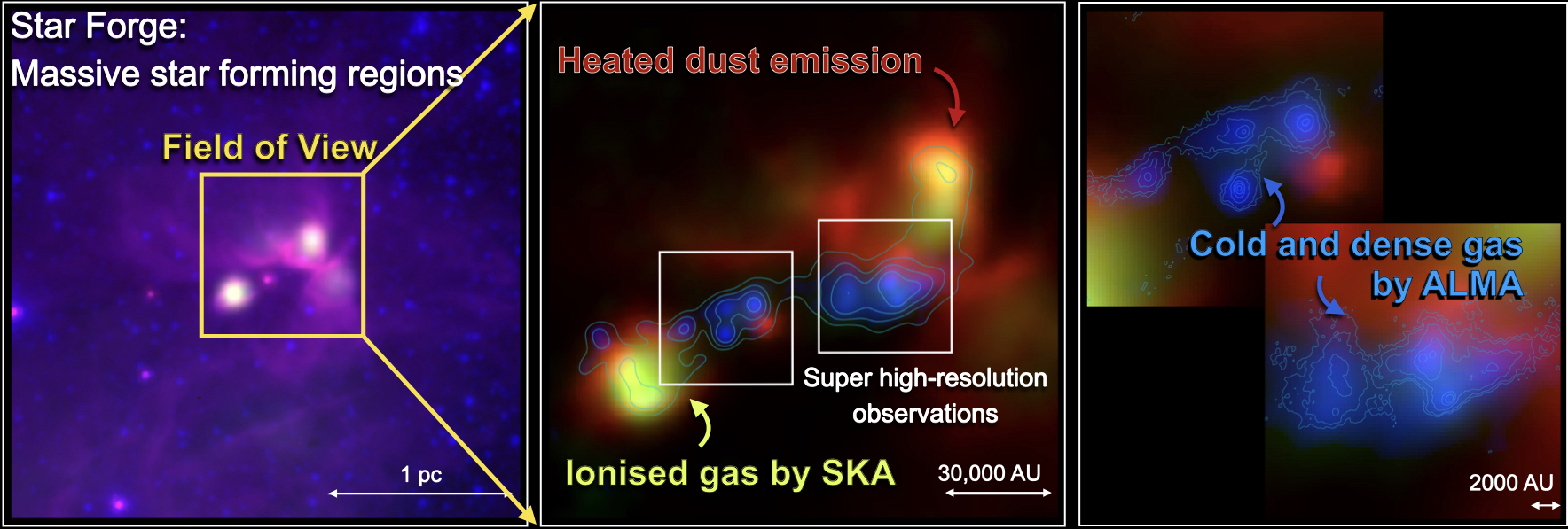}
    \caption{The concept of synergy between SKA-Mid and ALMA GP surveys. \textit{Left panel}: the Spitzer 3.6\,$\mu$m and 8\,$\mu$m and MeerKAT 1.28\,GHz composite image shows the overview of a massive star-forming region. \textit{Center panel}: The ionized gas is traced by the MeerKAT 1.28\,GHz and ALMA 3-mm images. \textit{Right panel}: With the help of ALMA-QUARKS data, we can even trace the cold and dense gas which are the site of new generation of star formation.}
    \label{fig:quarks}
\end{figure}

However, we are still not able to predict how much the environment and the feedback mechanisms contribute to the formation of one, or many fragments within each star forming regions. As shown in an example in Figure~\ref{fig:quarks}, all these mm surveys mainly traces the cold dusty envelopes of the star-forming regions. Only the synergy between ALMA, NOEMA and radio surveys will lead us to a comprehensive understanding on the interplay between gravity, turbulence and feedback at various scales, thanks to the mapping of the ionized gas that traces newly formed massive protostars, which are likely to be the main source of feedback in these star forming regions.

%%%%%%%%%%%%%%%%%%%%%%%%%%%%%%%%%%%%%%%%%%%%%%%%%%%%%%%%%%%%%%%%%%%%%%%%%%%%%%%%%%
\subsection{Low-Frequency Radio Continuum Surveys (\texorpdfstring{\boldmath$\nu \lesssim 8$\,GHz}))}
\label{section_RadioContinuumSurveys}

A number of GP radio continuum surveys have been carried out below 5 GHz. The most relevant in terms of sensitivity and angular resolution are: MAGPIS ($\simeq1.4$ and 5 GHz), CORNISH (5 GHz), THOR (1–2 GHz), GLOSTAR (4–8 GHz) and SMGPS (L-band). These surveys have delivered extensive catalogs of compact and extended radio sources, including \hii regions, SNRs, and PNe, with increasingly improved angular resolution and sensitivity.

The  Multi-Array Galactic Plane Imaging Survey (MAGPIS) were acquired in about 22 years of Karl G. Jansky Very Large Array (VLA) observations in seven out of the eight possible configurations \citep{White05}. The 1.4\,GHz survey covered $\simeq331$ deg$^{2}$ in the Galactic longitude range $-20^\circ < \ell < 120^\circ$ (and variable latitude range between $\pm0.8^{\circ}$ and $\pm2.7^{\circ}$) with a 90\% sensitivity completeness level of $\sim14$ mJy. The 5\,GHz survey mapped the GP region $-10^\circ < \ell < 42^\circ$, $|b| \leq 0.4^\circ$ with a 90\% sensitivity completeness level of $\sim2.9$ mJy. These programs have identified $\sim5000$ compact sources at 1.4 GHz and $\sim$2700 sources at 5 GHz. Furthermore, an extension of the 1.4\,GHz survey covered a portion of the first Galactic quadrant ($5^{\circ} < \ell < 48.5^{\circ}$), achieving an improved sensitivity of $\sim1$--$2$\,mJy using the VLA in B, C, and D configurations, which led to the detection of over 3000 compact sources and approximately $400$ extended sources, including $\sim50$ candidate supernova remnants \citep{Helfand06}.

CORNISH is a 5~GHz survey of the northern region of the inner GP ($295^\circ < \ell < 350^\circ$, $10^\circ < \ell < 65^\circ$, $|b| \leq 1^\circ$). The northern part of the survey ($10^\circ < \ell < 65^\circ$) was carried out using the VLA in B and BnA configurations \citep{Hoare2012,Purcell2013}. This yielded an angular resolution of $\sim 1.5''$ and a $1\sigma$ sensitivity better than 0.4~mJy~beam$^{-1}$. The survey was optimized to detect emission on scales smaller than $14''$, targeting the earliest phases of \hii\ region evolution. The southern part of the survey used the Australia Telescope Compact Array (ATCA) in its 6A configuration \citep{Irabor2023}. The ATCA observations provided an angular resolution of $2.5''$ and a $1\sigma$ sensitivity of 0.11~mJy~beam$^{-1}$. The CORNISH-North catalog, that is 90\% complete at a level of 3.9 mJy, detected $\sim2600$ reliable sources above a 7$\sigma$ detection limit, of which 288 and 170 are classified as \hii regions and planetary nebulae (PNe) respectively. The CORNISH-South survey detected a total of $\sim4700$ sources above $7\sigma$. In total, these surveys have classified $\sim800$ \hii regions and $\sim450$ PNe.

The HI/OH/Recombination line survey of the Milky Way (THOR; \citealt{Beuther2016}) and the Global View of Star Formation in the Milky Way (GLOSTAR; \citealt{Brunthaler2021}) used the upgraded WIDAR correlator of the VLA. In addition to providing a much larger continuum bandwidth, the surveys employed ``zoom'' modes to study selected spectral lines such as maser lines of OH and methanol, formaldehyde and various radio recombination lines (RRLs). The wide continuum bandwidth also enabled determination of the spectral index for bright sources. The THOR survey covered the GP with $14.5^\circ \leq l \leq 67.4^\circ$ and $|b| \leq 1.25^\circ$ from 1$-$2 GHz with the VLA in its C-configuration, yielding an angular resolution of 10-20\arcsec\ depending on the frequency. The data products include continuum images and data cubes for HI, the four OH maser lines and stacked hydrogen recombination lines. More than 10000 continuum sources have been classified as reliable detections \citep{Wang2018} including hundreds of \hii regions, SNRs and 164 PNe in the Milky Way. THOR continuum data have been also used to identify new Galactic SNR candidates \citep{anderson17}, whereas the line data (RRLs and OH) have been used to study massive star feedback \citep{rugel2019, rugel2025}.

The GLOSTAR survey covers the GP in the range $-2^\circ < l < 60^\circ$, $|b| < 1^\circ$ in addition to the Cygnus-X complex at a range $76^\circ < l < 83^\circ$, $-1^\circ < b < 2^\circ$ from 4$-$8~GHz. The unique aspect of the GLOSTAR survey is its utilization of data from interferometers in multiple configurations to achieve both good angular resolution and sensitivity to emission on large angular scales. The VLA was used in the D and B configurations, providing angular resolutions of $\sim 18''$ and $\sim 1.5''$, respectively. The GLOSTAR survey included observations from the 100-m Effelsberg telescope to obtain zero spacing information. The final data products include radio continuum images from D- and B-configurations individually, combined D- and B-configuration images, Effelsberg images, and those from combining Effelsberg and VLA D-configuration data. In addition, spectral line data for formaldehyde at 4.8~GHz, methanol masers at 6.7\,GHz, and hydrogen RRLs (six lines stacked together) are also available. Using the B-configuration GLOSTAR data, \citet{Dzib2023} and \citet{Yang2023} identified more than 300 \hii region candidates, including $\sim120$ new ones and $\sim40$ variable UC\,\hii regions (\citealt{Yang2025}). Using the D-configuration GLOSTAR data, hundreds of large-scale structures, large-scale structures such as star forming complexes and supernova remnants were also identified \citep{Medina2019,Medina2024}. Measuring the spectral index by combining the GLOSTAR and THOR surveys and comparing the radio continuum emission with the emission properties at other wavelength regimes from far to mid-infrared, these sruveys together allowed the identification and classification of more than 700 \hii regions. As with the THOR survey, polarization data have been used to identify SNRs (\citealt{Dokara2021,Dokara2023}).  In addition, the RRL data have been used to study the properties of 244 individual Galactic \hii regions in addition to studying the Galactic electron temperature gradient that arises from the gradient in the chemical composition of the Galactic disk \citep{Khan2024}.

Finally, the recently released SARAO MeerKAT GP survey (SMGPS; \citealt{Goedhart2024}) has mapped a large portion of the visible GP ($2^\circ < \ell < 60^\circ$, $252^\circ < \ell < 358^\circ$, $|b| \leq 1.5^\circ$) in L-band ($\sim$0.89-1.68\,GHz). With an angular resolution of 8\arcsec\ and a theoretical sensitivity of $\simeq10-20\ \mu$Jy/beam this survey is the largest, most sensitive and highest angular resolution survey of the GP at $\sim1$ GHz to date. From this survey it has been delivered a first catalogue of extended radio sources which includes $\sim4000$ known objects such as \hii regions, planetary nebulae, supernovae remnants, luminous blue variables and Wolf-Rayet stars \citep{Bordiu2025a}. A complementary compact source catalogue containing almost $\sim500\,000$ objects of the same region has recently become available (Mutale et al. 2025; in press).

\begin{figure}[!ht]
\begin{center}    
    \includegraphics[width=10cm]{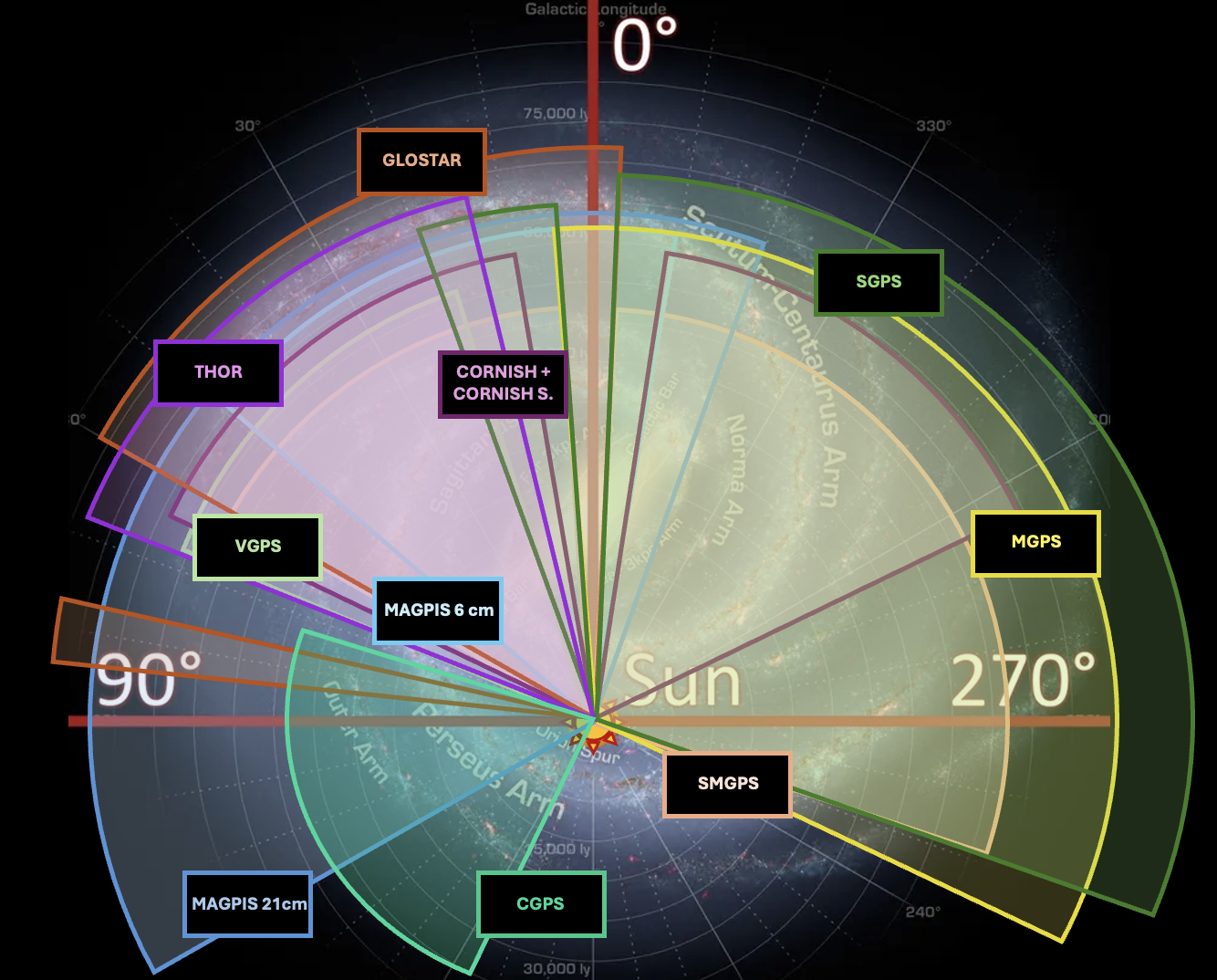}
    \caption{Spatial coverage of the main GP radio surveys in the frequency range $0.8 \lesssim \nu \lesssim 8$ GHz. The RACS survey is not displayed, as it provides full-sky coverage. The detailed observational parameters of each survey are summarized in Table \ref{tab:gpsurveys}.}
    \label{fig:radio_surveys}
\end{center}
\end{figure}

In Table \ref{tab:gpsurveys} we summarise the main parameters of these radio surveys and we include also for completeness the main observational parameters of other radio surveys of the GP at lower resolution and sensitivity, including the surveys performed with the Australian SKA pathfinder, ASKAP (adapted from \citealt{Bordiu2025a}). In Figure \ref{fig:radio_surveys} we visualize the GP coverage of all the radio surveys performed to date as described in Table \ref{tab:gpsurveys}.

\begin{table*}
\centering%
%\begin{threeparttable}
\scriptsize%
\caption{Main parameters of radio continuum surveys covering the GP in the frequency range 0.8--8.0 GHz. Adapted from \cite{Bordiu2025a}.}

\resizebox{\textwidth}{!}{
\begin{tabular}{lllllcccccl}
\hline%
\hline%
%Survey & Instr. & $l$ Coverage ($^\circ$) & $b$ Coverage ($^\circ$) & Gal. Quadrant & Freq. (GHz) & Bandwidth (MHz) & Ang. Resolution (") & rms ($\mu$Jy/beam) & LAS (') & Refs\\%
Survey & Instr. & $\ell$ Coverage & $b$ Coverage & Quadrant & Freq. & Bandwidth & FWHM & rms & LAS & Ref.\\%
& & (deg) & (deg) & & (GHz) & (MHz) & (arcsec) & ($\mu$Jy/beam) & (arcmin) & \\%
\hline%
{MAGPIS ($1.4$ GHz)} & {VLA} & -10$<l<$42 & $|b|<$0.4 & I, IV & {5.0} & {50} & {6} & {179} & {--} & {1}\\%
%& & 350<l<360 & $|b|<$0.4 & IV &\\%
\hline%
\multirow{4}{*}{MAGPIS ($5$ GHz)} & VLA+Effelsberg & 5$<l<$48.5 & $|b|<$0.8 & I & 1.4 & 95 & 6 & 897 & --\tablefootmark{a} & 2\\%
\cline{2-11}%
& \multirow{3}{*}{VLA} & $-$20$<l<$120 & $|b|<$0.8 & I, IV & \multirow{3}{*}{1.4} & \multirow{3}{*}{--\tablefootmark{b}} & \multirow{3}{*}{6} & \multirow{3}{*}{897} & \multirow{3}{*}{--} & \multirow{3}{*}{1}\\%
& & $-$10$<l<$40 & $|b|<$1.7 & I, IV & & & & & & \\%
& & 100$<l<$105 & $|b|<$2.2$^{\circ}$ & I & & & & & & \\%
\hline%
CORNISH & VLA & 10$<l<$65 & $|b|<$1.1 & I & 5.0 & 25 & 1.5 & 400 & 2 & 3\\%
\hline%
CORNISH-South & ATCA & 295$<l<$350 & $|b|<$1 & IV & 5.5 & 2000 & 2.5 & 110 & -- & 4\\
\hline%
GLOSTAR & VLA & $-2 <l< 60$ & $|b|<$1 & I & 4$-$8 & $2 \times 1024$\tablefootmark{c} & 1.5,18\tablefootmark{c} & 123 & 4 & 5\\
& & $76 < l < 83$ & $-1 < b < 2$ & & & & & & & \\%
\hline%
\multirow{2}{*}{THOR} & \multirow{2}{*}{VLA} & \multirow{2}{*}{14$<l<$67.4} & \multirow{2}{*}{$|b|<$1.25} & \multirow{2}{*}{I} & \multirow{2}{*}{1.42} & \multirow{2}{*}{128} & 18.1$\times$11.1 to & \multirow{2}{*}{300$-$1000} & \multirow{2}{*}{2} & \multirow{2}{*}{6}\\%
 & &  &  &  &  &  & 12.0$\times$11.6 &  & \\%
\hline%
\multirow{2}{*}{SMGPS} & \multirow{2}{*}{MeerKAT} & 2$<l<$60 & $|b|<$1.5 & I & \multirow{2}{*}{1.284} & \multirow{2}{*}{792} & \multirow{2}{*}{8} & \multirow{2}{*}{30} & \multirow{2}{*}{27} & \multirow{2}{*}{7}\\%
& & 252$<l<$358 & $|b|<$1.5 & III, IV\\%
\hline%
\\
\hline%
MGPS & MOST & 245$<l<$365 & $|b|<$10 & III, IV & 0.843 & 3 & 45$\times$45 csc$|\delta|$ & 1000 & 25 & 8\\%
\hline%
RACS-low{\tablefootmark{d}} & \multirow{2}{*}{ASKAP} & \multicolumn{2}{c}{$\delta \leq 41$ (34,240 deg$^{2})$} & \multirow{2}{*}{full sky} & 0.887 & 288 & 15-25 & 200-400 & $10\sim 30$ & 9\\%
RACS-mid & & \multicolumn{2}{c}{$\delta \leq 49$ (36,449 deg$^{2})$} & & 1.367 & 144 & $\geq8$ & 150-400 & $\sim5$ & 10\\%
\hline%
\multirow{3}{*}{VGPS} & \multirow{3}{*}{VLA+GBT} & 18$<l<$46 & $|b|<$1.3 & \multirow{3}{*}{I} & \multirow{3}{*}{1.4} & \multirow{3}{*}{1.866} & \multirow{3}{*}{60} & \multirow{3}{*}{--\tablefootmark{e}} & \multirow{3}{*}{--\tablefootmark{f}} & \multirow{3}{*}{11}\\%
& & 46$<l<$59 & $|b|<$1.9 & \\%
& & 59$<l<$67 & $|b|<$2.3 & \\%
\hline%
\multirow{2}{*}{SGPS} & \multirow{2}{*}{ATCA + Parkes} & 253$<l<$358 & $|b|<$1.5 & III, IV & \multirow{2}{*}{1.4} & \multirow{2}{*}{128} & 132 & \multirow{2}{*}{<1000} & \multirow{2}{*}{--\tablefootmark{g}} & \multirow{2}{*}{12}\\%
& & 5$<l<$20 & $|b|<$1.5 & I & & & 198\\%
\hline%
CGPS & DRAO Synth. Tel. & 74.2$<l<$147.3 & $-$3.6$<b<$5.6 & I, II & 1.42 & 35 & 60$\times$60 csc$|\delta|$ & 300 & 40 & 13\\%
\hline%
\end{tabular}
}
\tablefoot{
\tablefoottext{a}{Without Effelsberg data, the quoted LAS of VLA D-configuration is $\sim$16 arcmin.}
\tablefoottext{b}{Bandwidth variable across different observations, from 40 to 200 MHz.}
\tablefoottext{c}{Final mosaic obtained by integrating 8 sub-band images formed from two 1-GHz wide bands centred at 4.7 and 6.9 GHz. Angular resolution is $18''$ and $1.5''$ for D- and B-configuration respectively.}
\tablefoottext{d}{First data release.}
\tablefoottext{e}{Noise level $\sim$0.3 K.}
\tablefoottext{f}{Without Green Bank Telescope data, the quoted LAS of VLA D-configuration is $\sim$16 arcmin.}
\tablefoottext{g}{Without Parkes data, the theoretical LAS of ATCA smallest baseline configuration is $\sim$23 arcmin.}
}
\tablebib{(1)~\citet{White05};
(2) \citet{Helfand06}; (3) \citet{Hoare2012}; (4) \citet{Irabor2023};
(5) \citet{Brunthaler2021}; (6) \citet{Beuther16};
(7) \citet{Goedhart2024}; (8) \citet{Murphy07}; (9) \citet{McConnell20}; (10) \citet{Duchesne23}; (11) \citet{Stil06}; (12) \citet{McClure-Griffiths05}; (13) \citet{Taylor03}.
}
\label{tab:gpsurveys}
%\end{threeparttable}
\end{table*}

\subsection{The need for a survey of the Galactic Plane at 10-15 GHz}
\label{section:needs_SKA_survey}
Despite the rich legacy of these ancillary surveys, critical gaps remain in our ability to trace the full continuum spectral energy distribution (SED) of ionised sources, and to distinguish between emission mechanisms. In particular: free-free emission from \hii regions becomes increasingly dominant at higher frequencies ($\gtrsim 10$ GHz), while synchrotron emission steeply declines. Ultra- and hyper-compact \hii regions, deeply embedded in dense gas, often become optically thin only above $\sim$10 GHz. These are crucial phases for studying the onset of massive star feedback and their role in disrupting natal environments. Planetary nebulae and evolved stars often show a mix of optically thin thermal emission and recombination lines; their detection and characterization benefit from higher-frequency sensitivity, especially when embedded in complex fields. For SNRs, while typically dominated by synchrotron emission, high-frequency observations enable better background discrimination and separation from overlapping \hii regions or PNe.

The spectral index ($\alpha$) across the radio SED provides a direct probe of source optical depth, geometry and internal physical structure. Measurements based solely on low-frequency surveys can misclassify compact sources or miss optically thick components. A dual-frequency SKA-Mid survey centred at $\sim10$ and $\sim15$ GHz is optimal for isolating the thermal component and it would provide two well-separated SED points, enabling robust spectral index estimation even for unresolved sources and improving classification of jets, winds, and ionised bubbles.

\begin{figure}[!ht]
\begin{center}    
    \includegraphics[width=10cm]{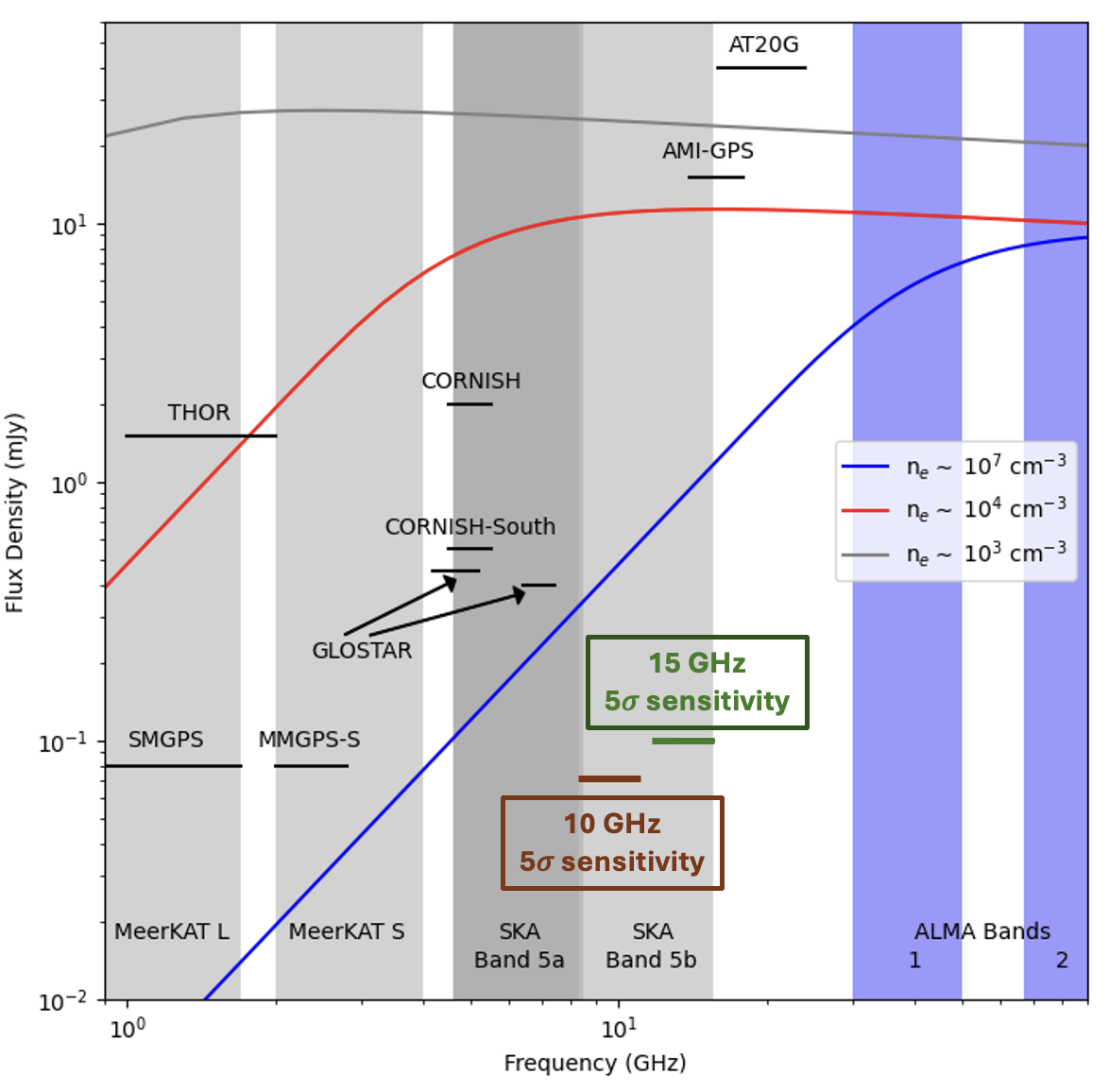}
    \caption{Sensitivity and frequency coverage of selected radio surveys, overlaid with the SEDs of \hii\ regions characterized by different average electron densities: $n_{e}\sim10^{7}$ cm$^{-3}$ (blue curve), $n_{e}\sim10^{4}$ cm$^{-3}$ (red curve), and $n_{e}\sim10^{3}$ cm$^{-3}$ (grey curve). The figure illustrates that the target sensitivity of the proposed GP survey at 10–15 GHz (20 $\mu$Jy at 15 GHz; see Section \ref{sec:technical_requirements}) will enable the detection of even the faintest \hii\ regions across the Galaxy with more than 5$\sigma$ sensitivity (100$\mu$Jy).}
    \label{fig:SED_HII}
    \end{center}
\end{figure}

Finally, none of the existing surveys provide uniform, wide-area coverage of the GP at frequencies above 8\,GHz. As showed in Figure\,\ref{fig:SED_HII}, a high-frequency radio continuum survey with arcsecond resolution and sub-mJy sensitivity would uniquely fill this gap, enabling a statistically robust, morphologically resolved census of ionised sources throughout the Galaxy. Its legacy value lies not only in expanding the known population of \hii regions, jets, SNRs, and PNe, but in providing the critical continuum anchor points needed to interpret the physics and evolution of these objects in synergy with molecular and FIR data.

%%%%%%%%%%%%%%%%%%%%%%%%%%%%%%%%%%%%%%%%%%%%%%%%%%%%%%%%%%%%%%%%%%%%%%%%%%%%%%%%%%
\section{The new science unveiled with a SKA-Mid GP survey at 10-15 GHz}\label{sec:science_with_GP}

In this Section we highlight more in detail all the new science that can be unveiled with observations of the GP at 10-15 GHz, with particular emphasis on the advantages of performing a GP survey in terms of statistics and coverage of different environments, ranging from \hii regions (Section \ref{section_Feedback_HIIRegions}), to radio jets (Section \ref{sec:radio_jets}), SNRs (Section \ref{sec:supernovae}), PNe (Section \ref{sec:planetary_nebulae}), and evolved massive stars (Section \ref{sec:evolved_massive_star}).

%%%%%%%%%%%%%%%%%%%%%%%%%%%%%%%%%%%%%%%%%%%%%%%%%%%%%%%%%%%%%%%%%%%%%%%%%%%%%%%%%%
\subsection{\texorpdfstring{\hii}) regions}
\label{section_Feedback_HIIRegions}

Massive O and B stars (M$_*$ $\geq$ 8 M$_{\odot}$) have a profound impact on their surrounding medium all along their life through their radiation and wind. During the formation process they produce copious amounts of ultraviolet photons ($E \geq 13.6$~eV) that ionize their surroundings, forming bubbles of hot ionized gas known as \hii regions. These regions expand rapidly, driving ionization fronts that launch shocks into the ambient medium. Recent JWST-MIRI images of nearby galaxies, such as that of the Phantom galaxy (NGC~628), show how important their impact is in shaping the surrounding molecular medium \citep{2023ApJ...944L..24W}. The effect of early radiative feedback on the distribution of molecular material has also been demonstrated in a larger sample of the nearby galaxies sampled by PHANGS, ALMA, and MUSE \citep{2023A&A...678A.171Z} on a spatial scale of 100~pc, providing important constraints on the efficiency of radiative feedback produced by intense and sudden bursts of star formation. This is indeed a critical parameter in galaxy evolution models \citep{Hopkins20}. However, the impact that such feedback might have on star formation at the resolved scales of few thousands of AU up to few pc is highly debated, despite dedicated observation programs such as FEEDBACK \citep{2020PASP..132j4301S}. Some numerical simulations suggest that this feedback is destructive \citep{2013MNRAS.435..917W,2015MNRAS.448.3248G}, but it is still quite debated. Other numerical simulations indicate that, under certain conditions, feedback may instead trigger star formation before the cloud dispersal \citep{Suin24}. Several observations support the triggering scenario, favouring the formation of a new generation of high-mass stars observed at the edges of ionized (H\,{\sc{ii}}) regions in the GP \citep{2010A&A...523A...6D, Thompson2012, 2017A&A...605A..35P, 2021A&A...646A..25Z, 2024ApJ...970L..40S}. More generally, the impact of feedback from \hii\ regions depends on several factors, including the strength of the magnetic field in the surrounding molecular cloud, as well as the spatial configuration and total mass of the ionising sources driving the feedback \citep{Suin25}.

The complex nature of this early radiative feedback and its influences requires deeper investigation and a statistically significant analysis of young massive star forming regions across various Galactic environments. Young \hii\ regions can be identified and characterised using a well-defined set of observational diagnostics.

The earliest phases are characterized by the formation of hypercompact (HC\,\hii), then ultracompact (UC\,\hii), and finally a classical \hii regions. Each phase is characterized by distinct physical conditions: HC\hii regions represent the very youngest phase of ionized gas, typically with diameters less than about 0.05 pc, high emission measures ($EM > 10^{10}$ pc cm$^{-6}$) and electron densities exceeding $10^6$ cm$^{-3}$ \citep{Kurtz05}. They are also characterized by extremely broad radio recombination lines (RRL), with typical $\Delta\ V= 40-50$ km s$^{-1}$, and in some cases up to $\simeq100$ km s$^{-1}$ \citep{Jaffe99, Sewilo04}.

As the ionized region expands, it quickly evolves into an UC\,\hii region, with typical scales of $\simeq0.1$ pc, electron densities of $10^{4}$--$10^{5}$ cm$^{-3}$, emission measures $EM \sim 10^{7}$--$10^{9}$ pc cm$^{-6}$ and $\Delta\ V= 25-30$ km s$^{-1}$ \citep{Wood1989}. UC\,\hii regions are often associated with masers and hot molecular cores \citep{Hoare2007}.

With time, they expand into larger, classical \hii regions—ionized zones extending up to several parsecs, with densities ($>100$ cm$^{-3}$), and visibly shaping their natal molecular clouds \citep{Anderson_WiseHII_2014ApJS}. 

Another crucial diagnostic of the different evolutionary phases is provided by the radio spectral energy distribution (SED), in particular the turnover, or ``knee,'' frequency where the spectrum transitions from being optically thick $(S_\nu \propto \nu^{2})$ to optically thin $S_\nu \propto \nu^{-0.1}$ free-free emission. This frequency is strongly dependent on the emission measure and electron density of the ionized gas, and it therefore shifts systematically as the \hii region evolves. In HC\,\hii regions, the combination of extremely high electron densities and emission measures pushes the turnover to very high frequencies, often $\geq20$ GHz (\citet{Kurtz05}, see also Figure \ref{fig:SED_HII}). As the ionized gas expands and densities decrease in the ultracompact phase, the knee frequency correspondingly shifts downward, typically into the $5-15$ GHz range, while still reflecting significant optical depth at centimeter wavelengths \citep{Yang2019}. In the more extended, evolved classical \hii regions, the densities are low enough that the emission becomes optically thin already at frequencies below $\sim1$ GHz, producing the nearly flat free--free spectra that characterize mature ionized nebulae \citep{Anderson_WiseHII_2014ApJS}. Although the evolution of \hii regions is more accurately described as a continuum rather than discrete phases, distinguishing different stages remains valuable for capturing key evolutionary moments, particularly the earliest phase of HC\,\hii region formation. This initial phase is critical, as the emerging ionization front begins to significantly influence the star formation process.

Nevertheless, a comprehensive and statistically significant study of HC\hii regions (along with a thorough radio characterization of UC\hii regions) remains lacking. This is primarily due to the rarity of these objects and the considerable challenges inherent in their identification. The extreme compactness of the HC\,\hii phase (size below $\simeq0.05$ pc), which requires angular resolutions better than $\simeq0.5$\arcsec\ to be resolved at distances up to 10 kpc, combined with the high optical depth at frequencies around 1–8 GHz, where most surveys have been conducted to date, and their relatively short lifetimes ($<10^{5}$ years; \citealt{Gonzalez-Aviles05, Sabatini21}), have made their detection and detailed study exceedingly difficult.

As a result, the identification of HC\hii regions has remained limited to $\sim50$, with most discoveries occurring serendipitously \citep{Yang2019, Yang2021}. However, this number has increased significantly in recent years by the Search for Clandestine Optically Thick Compact \hii Regions (SCOTCH) program \citep{Patel23,Patel24,Patel25}. This program systematically mapped the positions of $\sim500$ class II methanol masers located in the fourth quadrant and between $2^{\circ} \leq \ell \leq 20^{\circ}$ longitude range. These methanol masers are embedded in dense, high-mass clumps (\citealt{urquhart2013_mmb, urquhart2015_mmb}) and are known to be exclusively associated with high-mass star formation (\citealt{Pandian2010,breen2013}). These sources were observed with the ATCA at frequencies up to 24 GHz and an angular resolution reaching 0.5\arcsec\ and have resulted in the identification of 33 new HC\,\hii regions, more than doubling the number previously known \citep{Patel25}. 

The slightly larger size and extended lifetime of UC\,\hii regions have allowed the identification of several hundreds of such objects. Their characterization primarily relies on observations at  $\simeq5.5$ GHz, with confirmed near- and mid-infrared counterparts, achieved at an angular resolution of 2.5\arcsec\ \citep{Irabor2023}, corresponding to $\simeq0.1$ pc at a distance of 10 kpc. These UC\,\hii regions are often physically associated with large-scale diffuse ionized emission, as demonstrated in recent GP radio surveys such as MAGPIS, THOR, and GLOSTAR (see Section \ref{section_RadioContinuumSurveys}). However, the exact mechanisms behind the formation of such structures are still unclear. A simulation by \cite{wood2005} indicates that these extended envelopes may have been produced by the leakage of ionizing photons selectively through low-density regions within the boundary of UC\,\hii regions. On the other hand, \cite{Williams2000} indicates that the hierarchical nature of molecular clouds may lead to the clumpy, non-uniform density structures seen in  UC\,\hii regions as ionizing photons permeate these clouds. A recent study \citep{Dey2024} involving UC\,\hii regions with extended emission reveals electron densities ($n_e$) are within the range of 40 to 260~cm$^{-3}$, with a median value of 97~cm$^{-3}$, which is much different than what has been measured in the "standard" warm ionized medium (WIM; $n_e$ $\sim$ 0.1~cm$^{-3}$) and typical UC \hii regions ($n_e$ $\sim$ $10^4$~cm$^{-3}$).  Similarly, in another study, \cite{Goldsmith2015} reported $n_e$ of 10--100~cm$^{-3}$ with a mean of 29~cm$^{-3}$ within the GP using [N\,{\sc{ii}}] fine structure line surveys. Numerous hypotheses are suggested in the literature to explain these intermediate $n_e$ values, including multiple density components within the WIM, and the extended low-density envelopes of \hii regions interacting with the large-scale ISM. However, the exact physical processes that could lead to these intermediate $n_e$ values remain elusive.

Although the number of identified HC\hii\ and UC\hii\ regions has become significant, these surveys lacked the sensitivity and angular resolution required to probe the far edges of the Galaxy and to uniformly sample a large population of massive star-forming clumps across different Galactic environments, where extremely young \hii regions may still remain undetected. As a result, neither a systematic and unbiased census of HC\hii regions on Galactic scales nor a statistically significant characterization of the radio SEDs of HC\hii, UC\hii, and \hii regions across a vast portion of the Galactic Plane has yet been carried out. As illustrated in Figure \ref{fig:SED_HII}, the proposed GP survey in band 5b will represent a transformative advancement in our understanding of the feedback mechanisms produced by ionised emission from recently formed \hii regions. This survey will uniquely cover the $\sim$10–15 GHz frequency range, essential for constructing the radio SEDs of \hii regions across various evolutionary stages. Thanks to the unique capabilities of SKAO in its AA4 configuration, this GP survey will have a resolution as high as $\simeq0.08$\arcsec, a factor of $\simeq10$ better than the actual radio surveys across candidate HC\,\hii regions, which will allow us to resolve HC\hii regions up to distances of 20 kpc away from us.
With the proposed sensitivity of $\simeq0.08\ \mu$Jy/beam (see Section\,\ref{sec:technical_requirements}) it will be possible to probe UC\hii regions down to electron densities of the order 850-3000 particles /cm$^{3}$ (at a distances of 1 and 12 kpc, respectively, following the calculations in \citealt{Schmiedeke16}), almost one order of magnitude lower than typical UC\,\hii regions densities \citep{Yang2021}.

More details about the role of SKAO in unveiling the physics of \hii regions and their feedback are presented in the chapter "The Impact and Environment of Massive Stars and Stellar
Clusters" \citep{LorenAnderson01.2026.SKA}.

%%%%%%%%%%%%%%%%%%%%%%%%%%%%%%%%%%%%%%%%%%%%%%%%%%%%%%%%%%%%%%%%%%%%%%%%%%%%%%%%%%
\subsection{Radio Jets from massive regions}\label{sec:radio_jets}
%% TMR: First draft/brain dump; needs discussion, references and editing.

In low-mass protostars, accretion is facilitated by the removal of angular momentum via highly collimated, fast-moving jets ($\sim$100~km~s$^{-1}$), which in turn generate molecular outflows, as well as through magneto-hydrodynamic (MHD) disc winds \citep{Frank2014,Tabone2020}. Investigations in this context \citep[e.g.][and references therein]{Skretas2023,Maud2015} reveal that properties such as outflow mass, momentum, and the entrainment rate of surrounding gas correlate with the luminosity of the star-forming region. In contrast to their low-mass counterparts, massive protostellar jets are typically located at greater distances and within highly complex, clustered environments \citep[e.g.,][]{Li20, Li22, Izumi24, Lin25}. %This complexity has restricted detailed studies to only a few well-characterized cases. 
Although decades of observational studies \citep[e.g.,][]{Purser2016,Li2019,Rosero2019,karska2025} and numerical simulations \citep[e.g.,][]{Grudic22,Oliva2023,Gardiner2024,Lebreuilly2024,Hennebelle2024,neralwar2024} have underscored the pivotal role of jets and outflows as feedback mechanisms in galactic regions of high-mass star formation \citep[e.g.][and references therein]{Beuther25}, their precise physical nature remains far from fully understood. For this reason, expanding the currently limited sample is essential for advancing our understanding of the complete physical process: from how these massive jets are launched and propagate into their surrounding environments to how they regulate star formation efficiency.

The proposed SKA-Mid GP survey in Band~5b ($\sim$10-15~GHz), with its unprecedented cm-continuum sensitivity of 20-30~$mu$Jy~beam$^{-1}$, will enable, for the first time, a systematic, sensitive, and unbiased census of ionized jets from massive protostars across the Galaxy. To estimate the impact of this sensitivity, we first consider typical radio jet properties. Literature reports that typical flux densities of radio jets and knots in HH-objects at cm-wavelengths ($\sim$10~GHz) are in the range of $\sim 0.1$ to a few mJy \citep[e.g.,][]{Anglada2018}. Given that the spectral index in this frequency range is usually flat (median $\sim 0.45$), we can assume these radio luminosities are representative for the SKA-Mid GP survey frequencies ($\sim$9.5~GHz to 14~GHz). We use the radio luminosity to bolometric luminosity ($L_{\rm R} - L_{\rm bol}$) correlation from \cite{Anglada2018}, reads as:

\begin{equation}
    \left(\frac{S_\nu d^2}{\rm mJy\:kpc^2}\right) = 10^{-1.90}\:\left(\frac{L_{\rm bol}}{L_\odot}\right)^{0.59}\:.
\end{equation}

Based on this linear correlation, we can derive the minimum detectable bolometric luminosity ($L_{\text{bol, min}}$) as a function of source distance ($d$), using a $3\sigma$ detection limit ($\approx 60\,\mu\text{Jy}$ for the survey):
\begin{equation}
    \log_{10} (L_{\text{bol, min}}) = 3.2 + 1.7 \log_{10}(0.06\: d^2)\:.\label{eq:Lbolmin}
\end{equation}

Equation~\ref{eq:Lbolmin} yields the following $L_{\text{bol}}$ estimates: at $d = 100 \text{ pc}$, the detection limit is $\sim 0.006 \text{L}_\odot$ (meaning virtually all YSOs across the full range of protostellar masses with associated radio-jets would be detectable), while YSOs with $L_{\text{bol}} \gtrsim 14 \text{ L}_\odot$ will be detected at $d = 1 \text{ kpc}$. In terms of angular resolution, the SKA-Mid GP survey, with a proposed beam size of 80~mas, corresponds to $\sim$10~au at 140~pc or $\sim$80~au at 1~kpc. Given that emission in radio-jets is dominated by the central optically thick region at these scales, we anticipate that most sources will remain unresolved at this survey resolution. For a comprehensive overview of this topic, readers may refer to the chapter ``Jets and outflows in young stellar objects with the $\text{SKAO}$'' \citep{Sabatini01.2026.SKA}.

%%%%%%%%%%%%%%%%%%%%%%%%%%%%%%%%%%%%%%%%%%%%%%%%%%%%%%%%%%%%%%%%%%%%%%%%%%%%%%%%%%
\subsection{SNRs}\label{sec:supernovae}
The emission from SNRs is dominated by synchrotron emission arising by their relativist electrons. The typical averaged power-law spectrum of such emission is $S(\nu)\propto\nu^\alpha$, with values of $\alpha$ in the range $-1.1\leq\alpha\leq0.0$ \citep{Green25}, although there is increasing evidence that the power spectrum of an SNR can vary within its interior due to different distribution of the electrons, with variations of $\alpha$ across each SNR \citep{Loru2024}. The synchrotron emission dominates the spectrum at frequencies below the ones provided by our GP survey at 10-15 GHz where, instead, the free-free emission is usually the dominant mechanism of emission. However, it is very likely that in some cases even at these radio frequencies the dominant radiation is the synchrotron one generated by relativistic electrons, as predicted from theoretical models, at least at the shock fronts of \hii regions \citep{Padovani+2019}. In addition, it is yet unclear if the spectral index of SNRs remains constant up to such high frequencies and, in particular, if some variations may be expected locally across specific electron distributions within a given SNR. This is because only few studies of SNRs have been carried out at radio frequencies $\gtrsim 10$\,GHz, and those that exist to date, and they are usually limited to single-dish telescopes.

The expected number of SNRs in our Galaxy is also still debated: despite being among the  brightest radio sources in the sky, there is a severe discrepancy between known and expected Galactic SNR numbers, likely due to observational problems related to a lack of sensitive radio data and confusion in the GP \citep[][]{Brogan06}.

\begin{figure}
    \centering
    \includegraphics[width=0.7\linewidth]{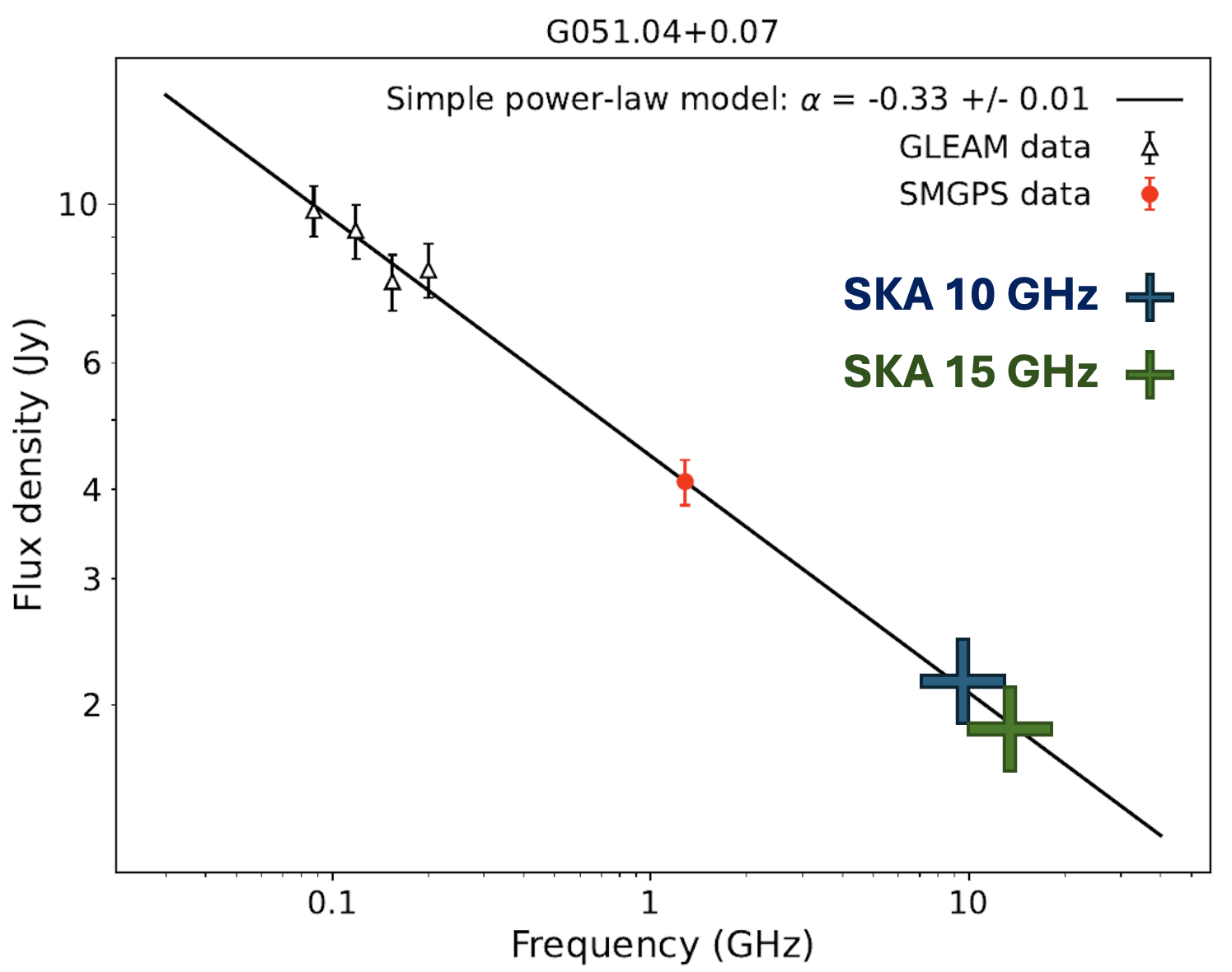}
    \caption{SED of the SNR G051.04+0.07 reported by \citet{Loru2024}. This is a typical example of a SNR for which only a few measurements are available, and high-frequency observations, in particular in the 10-15 GHz will give strong constraints.}
    \label{fig:g51}
\end{figure}

A high-frequency radio survey of SNRs would be invaluable both for identifying new SNR candidates and for conducting detailed spectral index analyses (see Figure~\ref{fig:g51}). This SKA survey of the GP has the potential to uncover hundreds of previously unknown Galactic SNRs. Currently, there exist over 300 SNR candidates \citep[e.g.,][]{anderson17, Dokara2021, anderson25}, primarily identified through observations with the JVLA and MeerKAT, which require spectral index measurements to confirm their nature as true SNRs. High-frequency radio observations will provide the key measurements to constrain the spectral energy distribution of both confirmed and candidate SNRs (Figure~\ref{fig:g51}). Moreover, the exceptional angular resolution achievable with SKAO in its final AA4 configuration will enable the resolution of individual SNRs up to distances of 20\,kpc, achieving a spatial resolution of approximately 1500\,AU  (see Section~\ref{sec:technical_requirements}). This capability will allow a detailed characterization of the local spatial variation in spectral indices across different regions of each SNR within the 10--15\,GHz frequency range. A comprehensive overview of the science that SKA can unveil for SNRs can be found in the chapter , readers may refer to the chapter "Supernova remnants in the new radio astronomy era" \citep{Ingallinera01.2026.SKA}

%%%%%%%%%%%%%%%%%%%%%%%%%%%%%%%%%%%%%%%%%%%%%%%%%%%%%%%%%%%%%%%%%%%%%%%%%%%%%%%%%%
\subsection{PNe}\label{sec:planetary_nebulae}
PNe represent one of the final evolutionary stages of stars with (ZAMS) masses between 0.08 and $8M_\odot$. After leaving the main sequence, these stars evolve first as red giants and then as AGB stars. In this phase, the stars eject part of their outer layers, which form a circumstellar envelope. After this phase, the stellar wind velocity increases by about an order of magnitude (while its density simultaneously decreases), and the star enters the post-AGB phase. In this stage, the star begins to expose increasingly hotter internal layers until what remains of the initial star reaches a surface temperature high enough to ionize the previously ejected circumstellar material, conventionally marking the birth of a PN.

PNe play a fundamental role in understanding stellar evolution, the chemistry of the interstellar medium, and even some fundamental physical processes related to magnetohydrodynamics. The final stages of the lives of these stars represent a critical moment when the products of nuclear reactions that occurred within them are expelled and alter the chemical composition of the interstellar medium. Understanding these phenomena provides general insights into cosmic abundances and, consequently, the formation of rocky bodies and even life. The origin of the various morphologies of PNe is still debated, but strong indications point to the influence of possible companion stars and interaction with magnetic fields.

For all these reasons, a systematic study of PNe is of fundamental importance, particularly of Galactic ones, which can be studied in greater detail. It is therefore necessary to have the largest possible sample. In fact, it is widely accepted that the number of known Galactic PNe (about 3000) is at least one order of magnitude less than expected. Estimates vary from just over 6000 to more than 40,000 depending on the method used to calculate the number (see, for example, \citealt{Sabin14}). This discrepancy likely stems, especially in the optical and IR, from the limited sensitivity of current surveys in the GP and from the fact that in these sky regions, the presence of dust absorbs radiation at short wavelengths and re-emits it in the mid- and far-infrared. In this context, radio observations can have a major impact in the search for the missing PNe across the Galaxy because they penetrate the dust, allowing us to observe regions inaccessible in other bands. This task will be fully addressed by future Galactic surveys conducted with SKAO. This potential has already been demonstrated by MeerKAT. Thanks to the SMGPS, 176 new PNe candidates have been identified by visual inspection. When compared with known PNe with the same angular sizes lying in the area covered by the survey, the new candidates show a distribution with a clear peak around Galactic latitude $b=0^\circ$, while the known PNe are approximately evenly distributed at $b<|1^\circ|$ \citep{Goedhart2024}. New radio surveys will hence further fill in the gap between known and expected PNe.

As the work with the SMGPS demonstrated, to assess the impact of SKAO, it is essential to consider not only sensitivity limits but also the ability to resolve sources: even if a PN is detected by these radio surveys, identifying it as such may be difficult if it appears point-like, as, for example, its SED might be confused with other source types. The exceptional resolution offered by SKA-Mid will then be fundamental to identifying new possible PN candidates where other probes (like H$_{\alpha}$) fail. Furthermore, the synergy with IR surveys such as GLIMPSE, MIPSGAL, WISE or Hi-GAL will enable a robust discrimination between radio emission associated with planetary nebulae and that arising from compact or bright \hii\ regions \citep{Hoare2012, Anderson12, Yang2023}.

In particular, the proposed survey at 10--15 GHz opens a new avenue for studying the whole population of PNe at global scales in the Galaxy. Above 5 GHz, PNe become optically thin, allowing meaningful morphological analyses and assessments of flux and variability. Assuming a standard spectral index of $-0.1$ for their thermal free-free spectra allows cross-frequency evaluations on both morphology and flux variability. The SKA-Mid survey will dramatically enhance our ability to track PNe over decades, and though few have attempted this, a striking example is the long-term radio monitoring of NGC 7027, where changes in flux density and stellar temperature over a 25-year period have been measured \citep{Zijlstra08}. These changes provided direct insight into the nebula's expansion and the evolution of its central star.

% The proposed survey at 10-15 GHz opens a new avenue for studying planetary nebulae PNe. Unfortunately, there have not been many extensive surveys at this frequency in the past and even less so in the southern hemisphere. However, it will still be possible to overlap with previous (mostly northern) 5 GHz surveys. Above 5 GHz, PNe become optically thin, permitting meaningful morphological analyses and assessments of flux and variability. Assuming a standard spectral index of -0.1 for their thermal free-free spectra allows cross-frequency evaluations on both morphology and flux variability in spite of frequency gaps. The SKA survey will dramatically enhance our ability to track PNe over decades, and though few have attempted this, a striking example is the long-term radio monitoring of NGC 7027, where changes in flux density and stellar temperature over a 25-year period have been measured \citep{Zijlstra08}. These changes provided direct insight into the nebula's expansion and the evolution of its central star.

%One clear enhancement in this SKA survey with respect to previous ones is the inclusion of radio recombination lines (RRLs), which are crucial in determining the values of electron temperature, density, and kinematics, even in dust-obscured regions. The most recent deep millimeter-wave RRL surveys covering both IC 418 and NGC 7027 (Huertas-Roldán et al 2025 arXiv:2506.23256) have demonstrated the diagnostic power of these lines.

%%%%%%%%%%%%%%%%%%%%%%%%%%%%%%%%%%%%%%%%%%%%%%%%%%%%%%%%%%%%%%%%%%%%%%%%%%%%%%%%%%
\subsection{Evolved massive stars}\label{sec:evolved_massive_star}

The evolution of massive stars remains a challenging topic in modern astrophysics, particularly concerning their late evolutionary phases \citep{Langer2012}. It is well established that massive stars undergo a series of short-lived stages before the final core collapse, characterised by widely different spectrophotometric features (e.g., Red Supergiants (RSGs), Luminous Blue Variables (LBVs), Wolf Rayet stars (WRs) and related objects). The occurrence and duration of these phases are influenced by several weakly constrained factors, such as initial stellar mass, rotation, and the mass-loss rate.

In the last decades, radio continuum observations have become an essential tool to investigate massive star mass-loss processes, being highly complementary to infrared observations. The potential of the radio window is threefold: first, observations at $\sim$GHz frequencies provide independent diagnostics of the current mass-loss rate \citep{Umana2005}, by means of the direct detection of thermal stellar winds, with a radio spectrum of the form $S_\nu \propto \nu^{0.6}$. Radio observations probe deeper within the wind than shorter wavelength observations; therefore, the resulting mass-loss rate estimates are less affected by the effects of wind structure and clumping. Second, radio observations allow for reconstructing the mass-loss history of the stars, through the detection of circumstellar shells or nebulae \citep{Umana2010,Buemi2017}. These structures, typically characterised by a flat radio spectrum ($\alpha\sim-0.1$), trace the extended ionised remnants of past mass-loss episodes. The characterisation of these structures contributes to more robust estimates of the total mass-loss budget over the lifetime of stars, which is a crucial piece of information for stellar evolution models. Last but not least, radio observations can also reveal more complex interaction phenomena, such as wind-wind collisions in massive binaries, characterised by non-thermal spectral indices, or bow shocks around runaway evolved massive stars, such as RSGs or WRs \citep{Eijnden2022}.

While targeted observations of single stars have advanced our understanding of these topics, population-sized studies have been hindered by the technical limitations of existing wide-area radio continuum surveys. However, SKAO precursors like ASKAP and MeerKAT are revolutionising the field, allowing for the first time for targeting  large samples of radio-emitting massive stars. The transformative potential of these facilities is already evident, with surveys like the SMGPS \citep{Goedhart2024} detecting a vast number of LBV and WR stars with a plethora of circumstellar structures around them (Umana et al., in prep., Buemi et al., in prep.), clear bow-shocks around cooler RSGs (Buemi et al., in prep.), and even allowing for the identification of new massive star candidates through the detection of circumstellar shells around unclassified variable stars \citep{Bordiu2025b}. Additionally, radio studies of massive stellar clusters are now possible at unprecedented angular resolution and sensitivity. Observing these coeval populations minimizes uncertainties in distance, metallicity, and formation conditions, providing a controlled environment to investigate the effects of the collective feedback from the cluster members.

SKAO will constitute an even larger leap forward, beyond the limitations of current facilities. The increased resolution will allow for disentangling the central stellar sources from the circumstellar envelope in the most compact objects, which appear as blended structures in current interferometric images; the increased instantaneous bandwidth will yield robust estimates of spectral indices, key to verifying the thermal or non-thermal nature of the emission of the stellar and nebular components; and the exceptional sensitivity will reveal faint, extended structures around known evolved massive stars and lead to the discovery of new candidates across the Milky Way. Specifically, a sensitive wide-area survey at 10-15 GHz would be particularly valuable; at these frequencies, the thermal emission from stellar winds is stronger, enabling the measurement of current-day mass-loss rates for nearly entire populations of evolved massive stars. Furthermore, it will allow us to directly monitor their variable radio emission, allowing for the study of the wind instabilities that characterise the final stages of massive star evolution.

The role of SKA in understanding the physics of evolved objects is detailed in the chapter "Evolved massive stars and their impact on their environment" \citep{Buemi01.2026.SKA}.

%%%%%%%%%%%%%%%%%%%%%%%%%%%%%%%%%%%%%%%%%%%%%%%%%%%%%%%%%%%%%%%%%%%%%%%%%%%%%%%%%%%%%%%%%%%%
\section{Insights from modelling and simulations}\label{sec:models_simulations}

%%%%%%%%%%%%%%%%%%%%%%%%%%%%%%%%%%%%%%%%%%%%%%%%%%%%%%%%%%%%%%%%%%%%%%%%%%%%%%%%%%%%%%%%%%%%
\subsection{Modelling free-free and synchrotron emissions}\label{sec:synchrotron_free_ree}
Radio emission from Galactic \hii regions comprises both thermal and non-thermal components. \cite{Meng+2019} analyzed VLA observations of Sgr B2(DS) between 4 and 12 GHz and identified  an expanding \hii region showing non-thermal emission all along the expanding shock. \cite{Padovani+2019} introduced a model attributing this emission to synchrotron radiation generated  by relativistic electrons accelerated at the shock front formed by the expansion of ionized gas.  Modelling such emission through direct comparison with radio spectra and morphology constrains key physical quantities,  including volume density, magnetic field strength, temperature, and flow velocity in the shock reference frame. The GLOSTAR survey confirmed the presence of non-thermal emission in both compact and extended \hii regions across the Galaxy \citep{Yang2023}, finding that a significant fraction of  sources exhibit negative spectral indices in the 4-8 GHz range. These results indicate that synchrotron radiation, although not the dominant emission mechanism, is emitted during the evolution of massive star-forming regions and is associated with the local acceleration of relativistic electrons and with the intensity of the magnetic field, which increases due to compression caused by the expanding shock.

At higher frequencies, such as the ones proposed for this survey, free-free emission by ionized gas is expected to dominate. Nevertheless, residual synchrotron contributions, especially in young or rapidly expanding \hii regions, may still influence observed spectra and polarization. Integrating synchrotron modelling into the analysis framework will enable a more comprehensive assessment of the thermal and non-thermal emission balance and facilitate the identification of regions where active particle acceleration and magnetic field amplification occur. 
Even when non-thermal emission is weak, establishing upper limits will refine constraints on magnetic field strengths, shock energetics, and the transition from ionized to molecular phases in massive star-forming complexes throughout the GP.

%%%%%%%%%%%%%%%%%%%%%%%%%%%%%%%%%%%%%%%%%%%%%%%%%%%%%%%%%%%%%%%%%%%%%%%%%%%%%%%%%%%%%%%%%%%%
%\subsection{Feedback mechanisms from H\,{\sc{ii}} regions in post-processed numerical simulations}\label{sec:RS_post_processing}
\subsection{Feedback mechanisms from \texorpdfstring{\hii}) regions in post-processed numerical simulations}\label{sec:RS_post_processing}

The classification of \hii regions based on their observed properties (Section \ref{section_Feedback_HIIRegions}) is further complicated by the tight coupling of the \hii region evolution with the properties of the local medium \citep{KK2001}. In fact, several HC\hii and UC\hii regions have shown rapid variations of a few percent per year in both luminosity and size \citep[][]{Franco-Hernandez2004, Rodriguez2007, Galvan-Madrid2008, Gomez2008, DePree2014}. However, it is unfeasible to obtain a detailed overview of the evolution of \hii regions and the interaction with their environments solely through observation.

% These regions do not leave in isolation, and the interation with the environment is such that recent studies even questioned the utility of a size-based classification of ultra and hyper compact \hii regions without accounting for the local environment \citep{Yang2019, Yang2021}.

Numerical simulations provide a powerful tool to investigate the temporal evolution of \hii regions and of their environmental conditions. In the last few decades, these studies have highlighted how the expansion of the ionised bubble can be delayed, confined, or even reversed. Local flows and turbulence \citep{Peters2010a, Peters2010b, Tanaka2016}, gravitational confinement \citep{Keto2003, Geen2015}, or pre-main sequence evolution \citep{Klassen2012b, Klassen2012a} may explain why some bubbles do not expand monotonically but exhibit a ``flickering'' behaviour instead \citep{Depree14}. Simulated regions showed good agreement with observational results, reproducing the observed flux variations of several percent per year \citep{Peters2010a}, and matching the lifetime of the ultra compact phase ($\gtrsim 10^5\,\mathrm{yr}$) derived from statistical samples \citep{Peters2010a, Klassen2012b, Geen2015}.

Nonetheless, significant uncertainties remain. Even with the most advanced simulations, it remains infeasible to simultaneously resolve the inner stellar scales of the assembling protostars (tens of AU and below) and track the larger-scale dynamics of the \hii region (from a few hundredths to tens of parsecs).
Therefore, the codes have to rely on subgrid models to schematically reproduce the formation of the massive star.
Furthermore, while some simulations achieve compact lifetimes consistent with observations, in some cases the compact phases lasted up to an order of magnitude longer \citep{Geen2015}. In such extreme environments, the uncertainties associated with the theoretical models are large, and different approaches may yield quantitatively different results \citep{Klassen2012b}.

This hinders the interpretation of the numerical results, and the physical origin of flickering is still debated.
It remains unclear whether the \hii regions variability is mainly driven by changes in the protostar itself \citep[][]{Franco-Hernandez2004, Klassen2012b} or by changes in the environment -- intermittent accretion flows \citep{Peters2010a}, local turbulence and outflows \citep{Tan2003, Tanaka2016}, or shielding from clumpy structures \citep[][]{Peters2010b}.

Therefore, tighter constraints are essential to disentangle the relative contributions of the various physical processes at play. Observations conducted at radio frequencies enable the examination of these processes through measurements of key local parameters such as electron density, ionisation structure, and gas kinematics. These diagnostics provide important benchmarks for numerical simulations, enabling direct comparison between model predictions and observed quantities.

% Therefore, tighter constraints are essential to disentangle the relative contributions of the various physical processes at play.
% As discussed above, observations at GHz frequencies provide a window to directly probe these earliest stages of high-mass star formation. Unlike optical or infrared observations, radio observations can pierce through the thick cocoons of gas and dust that surround forming massive stars. Moreover, the physical processes sampled through these frequencies (free-free emission and RRL \textcolor{olive}{-- This survey is only continuum, but perhaps we can link this to other proposed surveys focusing on RRLs}) enable the study of local properties, such as electron density, ionisation structure, and gas kinematics. These diagnostics provide important benchmarks for numerical simulations, enabling direct comparison between model predictions and observed quantities.

\begin{figure}
\begin{center}    
    \centering
    \includegraphics[width=15cm]{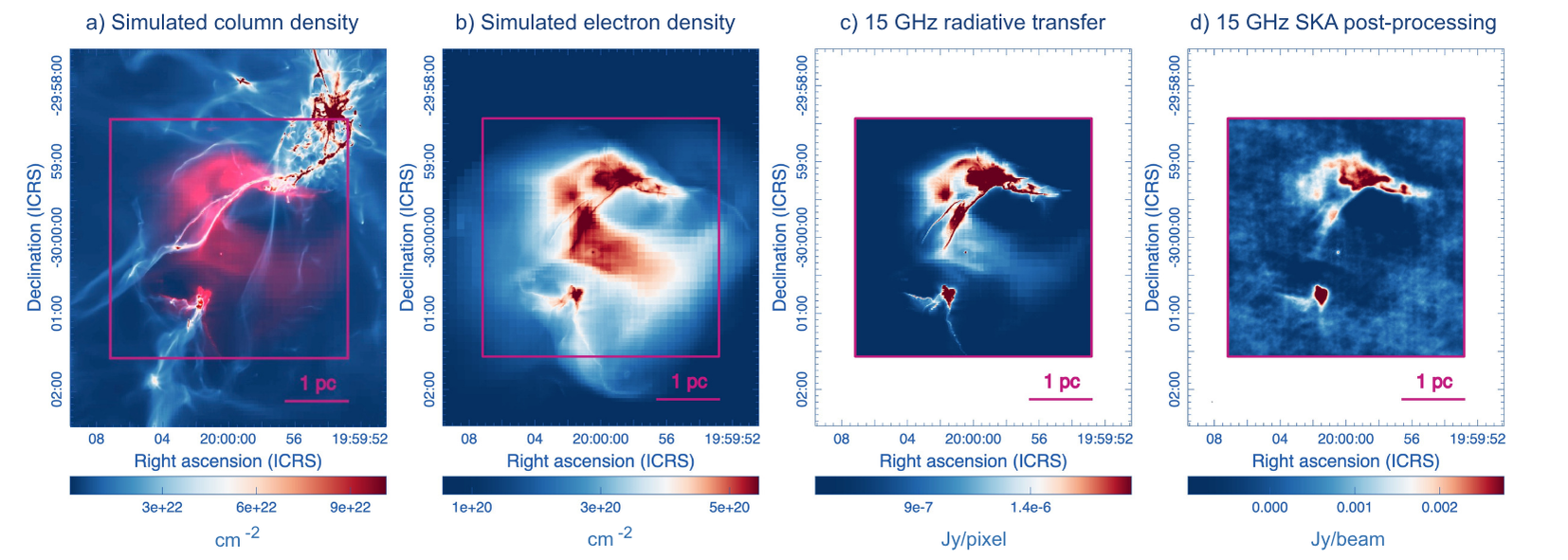}
    \caption{Rosetta Stone step-by-step production of SKA-Mid synthetic observations. \textit{Panel a}: Simulated column density ($N_H$) obtained with RAMSES for a collapsing 10$^4$\,M$_\odot$ cloud. Superimposed in magenta is the 15\,GHz emission. \textit{Panel b}: Simulated electron density ($n_e$) obtained with RAMSES. \textit{Panel c}: Radiative transfer at 15\,GHz performed with POLARIS. \textit{Panel d}: SKA-Mid-like post-processing obtained with a tailored CASA routine.}
    \label{fig:full_post_processing_SKA}
    \end{center}
\end{figure}

As recently demonstrated with the Rosetta Stone project (\citealt{Lebreuilly25, Tung25, Nucara25}), such comparison may be misleading, unless the simulations are post-processed in order to reproduce the observational patterns with the highest level of fidelity. To this end, and in preparation for this future survey of the GP with SKA-Mid, we are developing the first complete end-to-end framework to post-process a suite of simulations as if it was observed with SKA-Mid at 10-15 GHz. For our test case, we considered three-dimensional magnetohydrodynamic (MHD) simulation of a 10$^{4}$ \msun\ collapsing cloud developed with the RAMSES code \citep{Teyssier02}. Such cloud is in global collapse, it forms several cores and the most massive one develops an \hii region after $3.65\,$Myr.

The simulation has been post-processed using a novel implementation of the \textsc{POLARIS} code \citep{Reissl16}, specifically developed for this study, now incorporating free-free emission (Reissl et al., in prep.). 

Following the Rosetta Stone approach, we have simulated the post-processed ideal sky with a pipeline designed in the CASA software \citep{McMullin04} to reproduce the SKA-Mid array configurations and the integration time of our survey (see Section \ref{sec:technical_requirements}). The final product presents with the highest level of confidence how the simulated star-forming region would be seen through the SKA-Mid at 10-15 GHz.  
The RS framework steps are presented in Figure \ref{fig:full_post_processing_SKA}. The time-step corresponding to $\sim$0.4\,Myr  after the formation of the \hii region, is shown in Figure \ref{fig:full_post_processing_SKA}\,(a) with its associated electron density in (b). In Figure \ref{fig:full_post_processing_SKA}\,(c) we show the POLARIS post-processed map simulating the sky at 15 GHz. In Figure \ref{fig:full_post_processing_SKA}\,(d) is the post-processed map as seen with SKAO using the configuration of our GP survey at 15 GHz, assuming a pixel size of $0.02$\arcsec\ and an integration time of 20 seconds per pointing (Section \ref{sec:technical_requirements}).

The synergy between SKAO observations and simulations will provide an accurate framework for exploring the first stages of the interaction between massive stars and their natal environments, and enable a robust test for the competing theories of massive star formation and \hii region evolution.

%%%%%%%%%%%%%%%%%%%%%%%%%%%%%%%%%%%%%%%%%%%%%%%%%%%%%%%%%%%%%%%%%%%%%%%%%%%%%%%%%%
\section{Synergies with other SKAO science cases}\label{sec:synergies}
In this Section, we provide a detailed discussion of the synergies between this survey and other proposed SKAO science cases. As already presented in Section \ref{sec:science_with_GP}, the data from this GP survey at 10-15 GHz will be instrumental in advancing our understanding of the physics underlying radio emission processes in diverse environments, such as \hii regions, radio jets, SNRs, PNe and evolved massive stars. Furthermore, this survey will serve as a extremely valuable complementary dataset in other science cases, as discussed in the following.

%%%%%%%%%%%%%%%%%%%%%%%%%%%%%%%%%%%%%%%%%%%%%%%%%%%%%%%%%%%%%%%%%%%%%%%%%%%%%%%%%%
\subsection{RRLs}\label{sec:RRLs}
Radio recombination lines (RRLs) provide complementary information on the physical properties of atomic/ionised gas originating from radio-bright continuum sources detected as part of the GP survey. They can be used to constrain the temperature and density of ionised gas \citep{Luisi2019}, the hardness of ultraviolet radiation \citep{Roshi2012}, and metallicity \citep{men22}. In addition, they may serve as a tracer of gas kinematics, contributing to the understanding of physical processes as well as determining the distances \citep{anderson09a}. 

As discussed in the Chapter "Spectroscopic surveys with the SKA probing the ionised and molecular Milky Way" \citep{Karska01.2026.SKA}, there are 17 H$\alpha$ RRLs alone available band 5b, accompanied by the weaker He and C RRLs. The predicted sensitivities of RRLs are sufficient to detect, among other sources, compact \hii regions at their earliest stages (see Section \ref{section_Feedback_HIIRegions}). While the pressure broadening might limit the utility of RRLs at lower frequencies, this is not expected to be the issue in Band 5b, where lines tracing lower density gas are located. Nevertheless, combining survey data at Band 5b with lower frequencies would provide a full census of \hii regions spanning the range of properties and evolutionary stages. 

%%%%%%%%%%%%%%%%%%%%%%%%%%%%%%%%%%%%%%%%%%%%%%%%%%%%%%%%%%%%%%%%%%%%%%%%%%%%%%%%%%
\subsection{Methanol masers}\label{sec:methanol_masers}
Within the $\sim$8-15\,GHz range of SKA-Mid band 5b, at least three known maser transitions are present, two methanol masers and one formaldehyde maser (see Table \ref{tab:masers}). These three transitions have been detected all towards star-forming regions. The most widespread one amongst these, and thus also best studied, is the 12.2 GHz \meth\ (methanol) maser. It is thought to be excited mainly by warm dust emission, closely connected to the YSO luminosity. In fact, the 12.2\,GHz methanol maser is found to be more sensitive to variability in the radiative pumping mechanism, making it an interesting tracer of accretion flares in high-mass stars \citep[see, for example, the case of NGC 6334I][]{MacLeod2018}.  \cite{breen2010} observed 113 6.7\,GHz methanol maser bearing sources with the Parkes 64-m telescope and detected 12.2\,GHz \meth\ towards 68 of these sources, out of which 30 were new detections.  The 9.9\,GHz class I \meth\ masers and the 14.4\,GHz \form\ masers are both relatively rare; both are potentially excited in shocked regions, such as where molecular outflows impact the ambient ISM or the expanding \hii region. 

Just as blind GP surveys have increased the number of known 6.7\,GHz \meth\ masers \citep[e.g.,][]{Pandian2007,green2009,breen2015,nguyen2022}, including the 12.2\,GHz line in the SKA-Mid GP survey will find new detections also for this transition. The complete characterization of these masers, together with information of ambient feedback guaranteed by the continuum survey at 10-15 GHz will help to understand better the initial stages of high-mass star formation. A complete description of the maser science enabled with SKA-Mid can be found in the dedicated chapter "Cosmic Rulers: Masers as Tools for Probing Structure in the Galaxy and Beyond, from au to kpc" \citep{Rygl01.2026.SKA}.

\begin{table}
    \centering
    \caption{The main maser transitions obesrvable in the $\sim10-15$ GHz spectral window.}
    \label{tab:masers}
    \begin{tabular}{l  l l l l }
    \hline
    Molecule & Transition  & Freq (GHz) & Occurence & Ref \\
    \hline
\meth\ & $9_{-1} - 8{-2}$ E($v_t$=0), class I & 9.936202 & low & \cite{slysh1993}\\
    \meth\ & $2_1 - 3_0$ E($v_t$=0), class II & 12.1786& high & \cite{Batrla1987}\\ 
    \form\ & $2_{1,1} - 2_{1,2}$ & 14.48848 & low& \cite{chen2017}\\ 
    \hline
\end{tabular}
\end{table}

%%%%%%%%%%%%%%%%%%%%%%%%%%%%%%%%%%%%%%%%%%%%%%%%%%%%%%%%%%%%%%%%%%%%%%%%%%%%%%%%%%
\subsection{The Galactic Center}\label{sec:Galactic_center}

The SKA-Mid GP survey will provide a complementary view of the Galactic Center  (GC) to the one provided by the SKAO GC survey (see the chapter "The nearest galactic nucleus: Studying the Galactic Centre with SKA MID", \citealt{Schoedel01.2026.SKA}). The SKAO GC survey will cover an area of 2.0$^\circ$$\times$0.4$^\circ$ (equivalent to 290 pc$\times$60 pc), centered
on the massive black hole Sagittarius A$^*$ (Sgr A$^*$) using SKA-Mid in Bands 2, 5a and 5b. This survey will perform 1-hour snapshots per pointing achieving continuum sensitivities of 5.3 $\mu$Jy/beam at 1.36 GHz, 0.7 $\mu$Jy/beam at 6.6 GHz, and 0.8 $\mu$Jy/beam at 11.9 GHz. Therefore, the SKAO GC survey will cover a smaller area than that proposed in this chapter (see below), but it will reach a higher sensitivity per pointing (by at least a factor of $\geq$4) compared to the 20 $\mu$Jy/beam sensitivity requested for the SKA-Mid GP survey. In addition, in the SKAO GC survey multi-epoch observations over a time-span of 10 years will be carried out toward the Nuclear Stellar Cluster (NSC) and comparison fields, to uncover the dark cusp around SgrA$^*$.  

Although with a more limited spatial extent than the SKA-Mid GP survey, the higher sensitivity of the SKAO GC survey will unveil the population of the faintest stellar remnants, massive stars and pulsars in the GC, and it will allow studies of the properties and physics of the large-scale magnetic field, the origin of non-thermal filaments (unveiling the morphology and distribution of the faintest ones) and of the Galactic chimney. The molecular line emission detected within the SKAO GC survey will also enable studies of the  present day star formation (not only in the Central Molecular Zone, but also at the intersection with the dust-lane), as well as the discovery of prebiotic molecules such as simple sugars \citep{Schoedel01.2026.SKA}. Note that the SKAO GC survey will not provide any intra band spectral index, as it will be provided by the SKAO GP survey, and hence the GP survey will provide complementary data on the SED of the objects that are detected by  both surveys.

%%%%%%%%%%%%%%%%%%%%%%%%%%%%%%%%%%%%%%%%%%%%%%%%%%%%%%%%%%%%%%%%%%%%%%%%%%%%%%%%%%
\section{The 10-15 GHz SKA-Mid GP survey design}\label{sec:technical_requirements}
We aim to map a large portion of the GP visible from the SKA-Mid site, in South Africa, for a total of $630$ deg$^{2}$ of the GP. This field of view includes the inner part of the I quadrant the II and the IV quadrant, including the Galactic Center ($-180^\circ < \ell < 30^\circ$, $\vert b\vert \leq 1.5^\circ$). 

Thanks to the SKAO correlator, we can simultaneously observe the GP around 10 and 15
GHz using two spectra windows with bandwidths of $\sim2.5$ GHz each. Since Band5B receivers will cover the range of frequencies $8.3\leq \nu\ \leq 15.3$ GHz, we will center the rest frequency of the two spectral windows to approximately 9.5 GHz and 14.0 GHz respectively. We will perform the calculation at 14.0\,GHz, which is the band with the smaller field of view and the longest integration time to reach the desired sensitivity. 

The SKAO AA4 standard configuration includes 133 antennas with a dish size of 15 m. According to the official SKAO specifications, at the reference frequency of 14 GHz, the AA4 configuration corresponds to a a field of view (FoV) of each pointing of $\sim6.0'$, a synthesized beam of $\simeq0.07$\arcsec\ (assuming natural weighting), and the largest recoverable scale is $\sim2'$. 

\begin{figure}
\begin{center}    
    \centering
    \includegraphics[width=15cm]{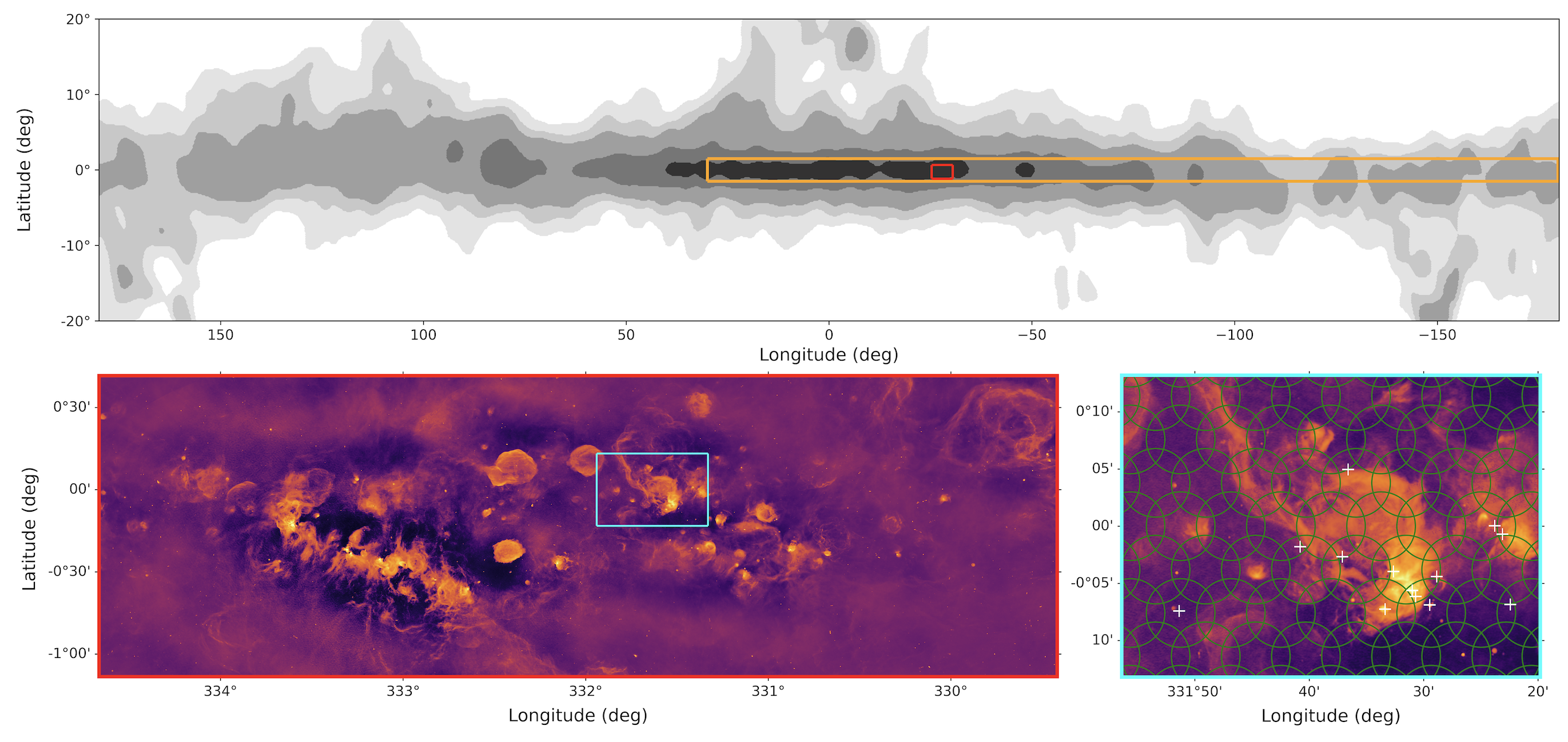}
    \caption{\textit{Top panel}: greyscale image of the Galaxy with overlaid the sky coverage of the SKA-Mid GP survey (orange box). \textit{Bottom left panel}: zoom-in on the Galactic region $330^\circ \lesssim \ell \lesssim 334^\circ$ observed at 1.3 GHz \citep{Goedhart2024}. This representative portion of the GP contains at least 3 confirmed supernova remnants, 4 additional SNR candidates, 5 PNe candidates, and $\sim80$ ALMAGAL sources. \textit{Bottom right panel}: a detail of the region with the sky coverage of the proposed mosaic survey at 15 GHz, with a FoV per pointing of $6'$.}
    \label{fig:survey_coverage}
    \end{center}
\end{figure}

With such FoV, in order to map the entire $630$ deg$^{2}$ of the proposed survey we would need in principle a total of $210^{\circ}\times 3^{\circ}/ (6.0'\times 6.0'/3600) = 63000$ pointings.  We also assumed a factor 1.96 due to the mosaicking to reach the desired sensitivity. This factor has been derived to guarantee the necessary degree of overlaps to reach uniform sensitivity across the entire GP. Combining all these parameters, we obtain a final number of $210^{\circ}\times 3^{\circ}/ (6.0' \times 6.0'/3600) \times 1.96 \sim 124000$ pointings. The sky coverage and survey strategy are showed in Figure \ref{fig:survey_coverage}.

The exquisite sensitivity of the SKAO telescope allows us to reach a sensitivity of 20 $\mu$Jy/beam with only $\sim20$ s of integration time per pointing at 14\,GHz (which is equivalent at $\sim14$ $\mu$Jy/beam per pointing at 10 GHz). This value does not take into account the increase in sensitivity achieved in mosaic mode due to the overlapping of the FoV of different pointings. We estimated this increase using the MeerKAT Mosaic calculator to look at the ratio of sensitivity achieved inside the area of a simulated mosaic with respect to the sensitivity achieved at the edge of the same simulated mosaic, where there is not the effect of the overlapping FoV. The ratio found with the MeerKAT Mosaic simulator is an increment in sensitivity of $\sim1.17$ thanks to the mosaicking. Therefore, we can reach the required sensitivity of 20 $\mu$Jy with a reduction of the integration time of a factor $(1/1.17)^{2}\simeq0.73$. The integration time therefore becomes $20\times0.73 \simeq 14.5$ s per pointing. Finally, we also consider a possible overhead factor of 20\%.

Altogether, the SKA-Mid survey of the GP at 10-15 GHz carried out in AA4 configuration requires a total tome of 14.5 s $\times 124000 \times 1.2 \simeq 600$ h.

The unparalleled combination of angular resolution and sensitivity achievable only with the SKAO in its AA4 configuration enables observations of sources extended up to 10 pc with a resolution of $\simeq1500$ AU at a distance of 20 kpc, sufficient to resolve the earliest stages of UC\hii region formation and to detect and characterize radio jets up to $\simeq10$ kpc away, while its sensitivity ensures the identification of even the faintest UC\hii regions with electron densities exceeding $10^{7}$ cm$^{-3}$ (Figure~\ref{fig:SED_HII}). Remarkably, all these results are achievable with only $\sim600$~h of observation time, making the proposed GP survey fully feasible.

If we assume the AA* array configuration, the time requested to complete the GP survey at the same sensitivity of $20\ \mu$Jy increases by a factor $\simeq3.1$, therefore the total time requested will be $\simeq1800$ h. At the same time, the synthetised beam will decrease from an average FWHM of $\simeq0.07$\arcsec\ to $\simeq0.16$\arcsec. Such survey will still be transformational for our understanding of the feedback mechanisms across the GP and it will allow to resolve UC\hii regions and radio jets up to $\simeq10$ kpc and $\simeq5$ kpc away from us, while missing the most distant objects within the GP.

Finally, we want to stress the importance to complement the SKA-Mid observations in band 5b with single-dish data. Every interferometer filters out spatial scales above a frequency-dependent threshold, the so-called Largest Angular Scale (LAS),  which can be approximated as $\theta_{\mathrm{LAS}}\sim\lambda/b_{\mathrm{min}}$, where $\lambda=c/\nu$ is the observation wavelength and $b_{\mathrm{min}}$ the interferometer's shortest baseline. Sources with an extension comparable to, or greater than the LAS are difficult to recover. From MeerKAT observations we can infer a LAS for SKA-Mid at $\sim\!1$~GHz of about half a degree \citep{Loru2024}. In band 5b, zero-spacing problems may arise for sources of the order of a few arcminutes. The combination of interferometer data with single-dish telescope can mitigate this limitation. A single-dish telescope with a diameter $D>b_{\mathrm{min}}$ can be used, to recover the missing information from the inner part of the $uv$-plane. In the southern hemisphere, the only suitable single-dish telescope is currently Parkes, which is planned to mount a new high-frequency receiver (UWH), expected in 2027, covering SKA-Mid bands 5A and 5B. Even if all-sky surveys with this new receiver are impractical, interesting portion of the sky may be imaged and combined with the data of the SKA-Mid GP survey in order to properly study sources more extended than the SKAO LAS.

%%%%%%%%%%%%%%%%%%%%%%%%%%%%%%%%%%%%%%%%%%%%%%%%%%%%%%%%%%%%%%%%%%%%%%%%%%%%%%%%%%
\section{Conclusions}\label{sec:conclusions}
The proposed 10–15 GHz SKA-Mid survey of the GP will represent a decisive step forward in our understanding of the ionised Milky Way. By bridging the gap between existing low-frequency radio surveys and the FIR and millimetre domain, it will provide the first comprehensive, high-resolution census of ionised structures, ranging from HC\hii to UC\hii regions, radio jets, PNe, SNRs, and evolved massive stars across the entire GP. This survey will take fully advantage of the spectacular sub-arcsec angular resolution, sub-mJy sensitivity, and wide-area coverage that SKAO in AA4 configuration can offer. With $\sim600$h of observations in Band 5b we will be able to map a total of $630$ deg$^{2}$ of the GP, including the Galactic Center ($-180^\circ < \ell < 30^\circ$, $\vert b\vert \leq 1.5^\circ$). Our survey will allow us to resolve feedback processes from massive stars and protostars over scales from a few hundred AU to several pc across a vast range of different Galactic environments.

The resulting dataset will constitute a legacy resource for decades to come, offering essential synergy with ALMA, MeerKAT, JWST, and future optical and IR surveys, and serving as the cornerstone for multi-scale analysis of star-formation feedback and its interplay with gravity and turbulence. 

Ultimately, the 10–15 GHz SKA-Mid GP survey will not only complete the continuum view of our Galaxy but will also establish a benchmark for future studies of ionised feedback in other galaxies, thereby advancing one of the central goals of modern astrophysics: to understand how stellar feedback governs the life cycle of baryons from the birth of stars to the assembly of galactic structure.

\section*{Acknowledgments}
\footnotesize{A.B. acknowledges financial support from the INAF initiative ``IAF Astronomy Fellowships in Italy'' (grant name MEGASKAT). OMS’s research is supported by the South African Research Chairs Initiative of the Department of Science, Technology and Innovation and the National Research Foundation (grant No. 81737). R.S. acknowledges financial support from the Severo Ochoa grant CEX2021-001131-S funded by MCIN/AEI/ 10.13039/501100011033 and from grant  PID2022-136640NB-C21 funded by MCIN/AEI 10.13039/501100011033 and by the European Union. A.Z. thanks the support of the Institut Universitaire de France. S.R., C.M., and F.M. acknowledge funding from the European Research Council in the ERC synergy grant “{\em ECOGAL} – Understanding our Galactic ecosystem: From the disk of the Milky Way to the formation sites of stars and planets” (project ID 855130). A.S-M.\ acknowledges support from the PID2023-146675NB grant funded by MCIN/AEI/10.13039/501100011033, and by the programme Unidad de Excelencia Mar\'{\i}a de Maeztu CEX2020-001058-M, as well as support from the RyC2021-032892-I grant funded by MCIN/AEI/10.13039/501100011033 and by the European Union `Next GenerationEU'/PRTR. G.S., Cl.Co. And L.P. acknowledge financial support under the National Recovery and Resilience Plan (NRRP), Mission 4, Component 2, Investment 1.1, Call for tender No. 104 published on 2.2.2022 by the Italian Ministry of University and Research (MUR), funded by the European Union – NextGenerationEU-Project Title 2022JC2Y93 Chemical Origins: linking the fossil composition of the Solar System with the chemistry of protoplanetary disks – CUP J53D23001600006 – Grant Assignment Decree No. 962 adopted on 30.06.2023 by the Italian Ministry of Ministry of University and Research (MUR); the project ASI-Astrobiologia 2023 MIGLIORA (“Modeling Chemical Complexity”, F83C23000800005); the INAF-GO 2024 fundings ICES, the INAF-GO 2023 fundings PROTOSKA (“Exploiting ALMA data to study planet forming disks: preparing the advent of SKA”, C13C23000770005); the INAF Mini-Grant 2022 “Chemical Origins” (PI: L. Podio) and the INAF Minigrant 2023 TRIESTE (“TRacing the chemIcal hEritage of our originS: from proTostars to planEts”; PI: G. Sabatini). E.B. acknowledges the support from the Italian Ministry for Universities and Research under the Italian Science Fund (FIS 2 Call - Ministerial Decree No. 1236 of 1 August 2023) and the Next Generation EU funds within the National Recovery and Resilience Plan (PNRR), Mission 4 - Education and Research, Component 2 - From Research to Business (M4C2), Investment Line 3.1 - Strengthening and creation of Research Infrastructures, Project IR0000034 – “STILES - Strengthening the Italian Leadership in ELT and SKA". M.P. acknowledges the INAF grant 2023 MERCATOR (``MultiwavelEngth signatuRes of Cosmic rAys in sTar-fOrming Regions'')
and the INAF grant 2024 ENERGIA (``ExploriNg low-Energy cosmic Rays throuGh theoretical InvestigAtions at INAF''). I.J-.S acknowledges funding from grant PID2022-136814NB-I00 funded by the Spanish Ministry of Science, Innovation and Universities/State Agency of Research MICIU/AEI/ 10.13039/501100011033 and by “ERDF/EU”, and from the ERC grant OPENS (project number 101125858) funded by the European Union. F.M., M.V-M. and N.C. acknowledge support from the from the French Agence Nationale de la Recherche (ANR) through the project COSMHIC (``Constraining the Origin of Stellar Masses in Hierarchical Infalling Clouds'', ANR-20-CE31-0009).}

\bibliographystyle{abbrvnat-maxbibnames4}
\bibliography{chapter.bib} % if your bibtex file is called example.bib

\end{document}